\documentclass[sigconf, nonacm]{acmart}

\newcommand\vldbdoi{XX.XX/XXX.XX}
\newcommand\vldbpages{XXX-XXX}
\newcommand\vldbvolume{14}
\newcommand\vldbissue{1}
\newcommand\vldbyear{2020}
\newcommand\vldbauthors{\authors}
\newcommand\vldbtitle{\shorttitle} 
\newcommand\vldbavailabilityurl{URL_TO_YOUR_ARTIFACTS}
\newcommand\vldbpagestyle{plain} 

\usepackage[normalem]{ulem}
\usepackage{booktabs}
\newcommand{\model}{InBox}

\begin{document}
\title{\model: Recommendation with Knowledge Graph using Interest Box Embedding}

\author{Zezhong Xu}
\affiliation{%
  \institution{Zhejiang University}
  \city{Hangzhou}
  \state{China}
}
\email{xuzezhong@zju.edu.cn}

\author{Yincen Qu}
\affiliation{%
  \institution{Alibaba Group}
  \city{Hangzhou}
  \state{China}
}
\email{yincen.qyc@alibaba-inc.com}

\author{Wen Zhang}
\affiliation{%
\institution{Zhejiang University\\Zhejiang University-Ant Group Joint Laboratory of Knowledge Graph}
  \city{Hangzhou}
  \state{China}
}
\email{Zhang.wen@zju.edu.cn}

\author{Lei Liang}
\affiliation{%
    \institution{Ant Group}
  \city{Hangzhou}
  \state{China}
}
\email{leywar.liang@antgroup.com}

\author{Huajun Chen}
\affiliation{%
\institution{Zhejiang University\\Alibaba-Zhejiang University Joint Institute of Frontier Technology\\Zhejiang University-Ant Group Joint Laboratory of Knowledge Graph}
  \city{City}
  \country{country}
}
\email{huajunsir@zju.edu.cn}

\begin{abstract}
Knowledge graphs (KGs) have become vitally important in modern recommender systems, effectively improving performance and interpretability. Fundamentally, recommender systems aim to identify user interests based on historical interactions and recommend suitable items. However, existing works overlook two key challenges: (1) an interest corresponds to a potentially large set of related items, and (2) the lack of explicit, fine-grained exploitation of KG information and interest connectivity.
This leads to an inability to reflect distinctions between entities and interests when modeling them in a single way. Additionally, the granularity of concepts in the knowledge graphs used for recommendations tends to be coarse, failing to match the fine-grained nature of user interests. This homogenization limits the precise exploitation of knowledge graph data and interest connectivity.
To address these limitations, we introduce a novel embedding-based model called {\model}. Specifically, various knowledge graph entities and relations are embedded as points or boxes, while user interests are modeled as boxes encompassing interaction history. Representing interests as boxes enables containing collections of item points related to that interest. We further propose that an interest comprises diverse basic concepts, and box intersection naturally supports concept combination.
Across three training steps, {\model} significantly outperforms state-of-the-art methods like HAKG and KGIN on recommendation tasks. Further analysis provides meaningful insights into the variable value of different KG data for recommendations. In summary, {\model} advances recommender systems through box-based interest and concept modeling for sophisticated knowledge graph exploitation.
\end{abstract}

\maketitle

\pagestyle{\vldbpagestyle}
\begingroup\small\noindent\raggedright\textbf{PVLDB Reference Format:}\\
\vldbauthors. \vldbtitle. PVLDB, \vldbvolume(\vldbissue): \vldbpages, \vldbyear.\\
\href{https://doi.org/\vldbdoi}{doi:\vldbdoi}
\endgroup
\begingroup
\renewcommand\thefootnote{}\footnote{\noindent
This work is licensed under the Creative Commons BY-NC-ND 4.0 International License. Visit \url{https://creativecommons.org/licenses/by-nc-nd/4.0/} to view a copy of this license. For any use beyond those covered by this license, obtain permission by emailing \href{mailto:info@vldb.org}{info@vldb.org}. Copyright is held by the owner/author(s). Publication rights licensed to the VLDB Endowment. \\
\raggedright Proceedings of the VLDB Endowment, Vol. \vldbvolume, No. \vldbissue\ %
ISSN 2150-8097. \\
\href{https://doi.org/\vldbdoi}{doi:\vldbdoi} \\
}\addtocounter{footnote}{-1}\endgroup

\ifdefempty{\vldbavailabilityurl}{}{
\vspace{.3cm}
\begingroup\small\noindent\raggedright\textbf{PVLDB Artifact Availability:}\\
The source code, data, and/or other artifacts have been made available at \url{\vldbavailabilityurl}.
\endgroup
}

\section{Introduction}
\begin{figure}[h]
\centering
  \includegraphics[scale=0.6]{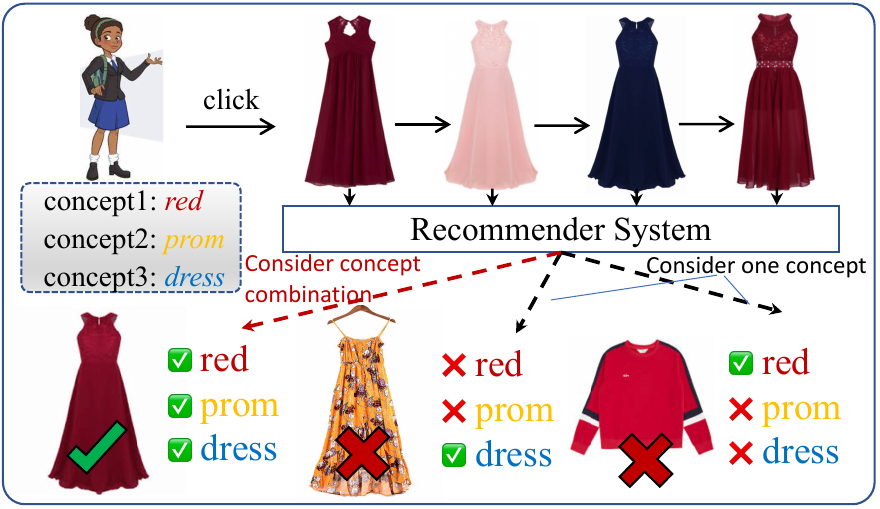}
  \caption{An example illustrating a specific interest is the combination of several concepts.}
  \label{fig:intro1}
\end{figure}
As the Internet develops, recommendation systems have emerged as vital tools for identifying user preferences across diverse domains, encompassing search engines, e-commerce platforms, and media services.
Conventional recommendation approaches~\cite{DLUI, CF, FM} primarily focus on collaborative filtering (CF). Although proficient at discerning collaborative patterns, these techniques grapple with issues of data sparsity and cold-start dilemmas.
In recent times, there has been a growing interest in incorporating knowledge graphs, replete with abundant entity and relational data, into recommender systems. Such integration not only enhances the precision of recommendation results but also addresses cold-start predicaments and bolsters the explainability of the recommendation pipeline.

Employing auxiliary information such as KGs revolves around procuring suitable representations for items and users.
Certain techniques investigate the utilization of knowledge graph pathways to bolster user-item interactions, thereby offering interpretability for recommendation outcomes. Nevertheless, these approaches necessitate labor-intensive feature engineering, such as manually crafted meta-paths, and grapple with subpar transferability and fluctuating performance.
Graph neural networks (GNN) have also gained traction in recommendation tasks. By implementing an information aggregation scheme, multi-hop neighbor data is amalgamated into node representations. Although effective, GNN-based models typically hinge on the quality of the knowledge graph; in real-world situations, KGs tend to be sparse and replete with noise, which curtails their efficacy.
Some of the early studies~\cite{CKE, HKB, DKN} incorporated item embeddings, derived from KG embedding (KGE) methodologies (e.g., TransE~\cite{transE} and TransR~\cite{transR}), as prior embeddings to augment the recommendation endeavor. While embedding-centric models profit from the straightforwardness and expressivity of KGE, they falter in capturing the higher-order dependencies between knowledge and user interests.

 Indeed, the crux of recommendation lies in discerning user preferences and suggesting suitable items accordingly (e.g., as manifested by user embeddings in certain studies). While previously discussed approaches yield commendable results within knowledge-aware contexts, we contend that they fall short in adequately representing user interests when taking the subsequent two factors into account.

Firstly, user interests typically encompass a potentially extensive collection of related items (e.g., a user fond of sports games may appreciate numerous titles such as \textit{Ring Fit Adventure}, \textit{Just Dance}, and \textit{Fitness Boxing}). Prior research often encapsulates user interests or the user itself as a single point in the embedding space, rendering the construction of effective connections between item sets and interests challenging, since mere points prove inadequate in delineating such one-to-many correspondences.
Furthermore, individual user interests can be exceptionally diverse and intricate, precluding the manual delineation of all conceivable preferences. We suggest that these interests may be discerned through the convergence of multiple fundamental concepts within KGs (e.g., in Figure~\ref{fig:intro1}, a girl may desire a \textit{crimson prom gown}, characterized by \textit{crimson}, \textit{prom}, and \textit{gown}; considering merely one or two of these facets may yield subpar recommendations). By consolidating concepts within KGs to construct the deterministic relationship between KG and interest definition, more accurate and personalized interests can be explicitly determined for distinct users. Unfortunately, existing models fall short of establishing a connection between the intersection of KG concepts and user preferences in a fine-grained way.
\begin{figure}[tbp]
\centering
  \includegraphics[scale=0.57]{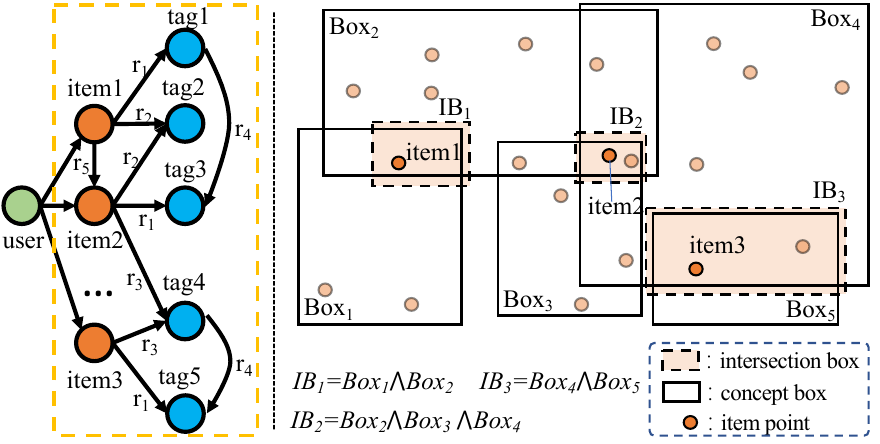}
  \caption{(a) An example of the knowledge-aware recommendation data. (b) An illustration of the item points and concept boxes.}
  \label{fig:intro2}
\end{figure}
To address the aforementioned challenges, we introduce a novel model, {\model}, wherein user interests are conceptualized as a box containing suitable items. KG entities are partitioned into \textbf{items}, those users have interacted with, and non-items, designated as \textbf{tags} in this paper, as illustrated in Figure~\ref{fig:intro2}. These two entity types are represented distinctly:
An item is embedded into the vector space as a point, whereas a tag or relation is denoted by a box instead of a single point. For instance, in a movie recommendation context, consider the triplet \textit{(Avatar, is directed by, James Francis Cameron)}: \textit{Avatar} is represented as a point since it is an item, while the relation \textit{is directed by} and the tag \textit{James Francis Cameron} are both depicted as boxes; the relation-tag pair can be interpreted as a concept (since a tag linked with different relations could express different meanings. Like \textit{(is directed by, James Francis Cameron)} and \textit{(is written by, James Francis Cameron)}), although the tags in are both \textit{James Francis Cameron}, these are two different concepts.
This approach offers several key advantages: Firstly, items connected to the same relation-tag pair could be modeled within the box region, to satisfy the one-to-many correspondences; Secondly, treating relation-tag pairs as concepts allows for the natural expression of concept combinations via box intersections, which means we could use the intersection region of two boxes as a combination of the two corresponding concepts; Finally, the intersection of two boxes remains a box (even an empty set can be considered a unique box), signifying that this logical operation over boxes is closed, thus facilitating the combination of any number of concepts.

In addition to defining the aforementioned representations, the model must possess the following distinct capabilities: 1) ensuring that the representations between items and tags satisfy their corresponding relationships within the KG; 2) defining the concept combination operation; 3) empowering the model to abstract user interests based on existing information.
To achieve this, the model training process consists of three distinct stages. The first stage focuses solely on the knowledge graph to obtain suitable representations that position item points within corresponding concept boxes. The second stage strives to situate items within the intersection box region of their associated concepts. The third stage employs interaction history to generate an interest box embedding, which is subsequently utilized to recommend potential items for users.
To accomplish these three stages, various distance functions are employed, encompassing point-to-point, box-to-box, and point-to-box distances.

To substantiate the efficacy of {\model}, experiments are executed on four real-world datasets. Experimental outcomes demonstrate that {\model} surpasses state-of-the-art models, including KGAT~\cite{KGAT}, KGIN~\cite{KGIN}, and HAKG~\cite{HAKG}, across all datasets. Additionally, model analysis reveals that different triplet types in KGs exert varying impacts on our model, potentially offering novel insights into the significance of KG triplets in KG-enhanced recommender systems.

In summary, our contributions to this work are as follows:
\begin{itemize}
\item We introduce a novel approach to modeling user interests for recommendation tasks, positing that interests can be represented by the conjunction of several general concepts in KGs.
\item We present {\model}, an innovative recommendation model that employs box embeddings to represent tags, relations, and user interests, fulfilling the criteria that an interest corresponds to a set of items and supports intersection logical operations.
\item Through experimentation, we prove that {\model} surpasses existing methods. Our model analysis unveils a fresh perspective for assessing the significance of various triplet types in knowledge-aware recommender systems.
\end{itemize}
\section{PROBLEM FORMULATION}
 In this section, we introduce the data in the KG-enhanced recommender system, which comprises the user-item interaction history and various triplet types in the knowledge graph, as well as the task definition.

\textbf{User-Item Interaction Graph.} Given a user set $\mathcal{U}$ and an item set $\mathcal{I}$, the user-item interaction graph can be constructed as $\mathcal{G}_{u} = {(u, i)| u \in \mathcal{U}, i \in \mathcal{I}}$, where each $(u, i)$ pair signifies that user $u$ has previously engaged with item $i$, as determined by implicit feedback~\cite{BPR} from $u$ to $i$.
Note In this paper, we do not differentiate between various user behaviors (e.g., clicks, purchases); the datasets and baselines employed in our experiments also refrain from making such distinctions. 

\textbf{Auxiliary Knowledge Graph.} KGs store structured information, including item attributes. Renowned KGs like Wordnet~\cite{Wordnet} and Freebase~\cite{Freebase} have been developed. Let $\mathcal{E}$ represent a set of entities and $\mathcal{R}$ a set of relations. A knowledge graph can be defined as $\mathcal{G}_{k} ={(e_1, r, e_2) | r \in \mathcal{R}, e_1, \in \mathcal{E}, e_2 \in \mathcal{E}}$, where each triplet $(e_1, r, e_2)$ indicates a relation $r$ connects entities $e_1$ and $e_2$.

In the recommendation context, the entity set comprises the item set $\mathcal{I}$ and other non-item entities, which we term tags in this paper, and these non-item entities can be considered as a tag set $\mathcal{T}$. Accordingly, all the triplets in $\mathcal{G}_{k}$ can be categorized into three groups: (1) $(item, relation, item)$ triplet, denoted as $IRI$ triplet; (2) $(tag, relation, tag)$ triplet, denoted as $TRT$ triplet; (3) $(item, relation, tag)$ triplet, denoted as $IRT$ triplet. Note that \textit{(tag, relation, item)} is regarded as the same type as $(item, relation, tag)$ only if we use the inverse relation, so we do not specifically distinguish this type. 

For instance, in the movie recommendation task, there are two items, $Avatar$ and $Avatar2$, and two tags, \textit{James Francis Cameron} and \textit{America}. The examples of different triplet types are as follows:
\begin{itemize}
\item \textit{(Avatar2, is sequel of, Avatar)} is an $IRI$ triplet.
\item \textit{(James Francis Cameron, is citizen of, America)} is a $TRT$ triplet.
\item \textit{(Avatar, is directed by, James Francis Cameron)} is an $IRT$ triplet.
\end{itemize}
The relation-tag pair, such as \textit{(is directed by, James Francis Cameron)}, can be viewed as a concept introduced earlier.
The three distinct types of triplets will be processed differently in the training step.

\textbf{Task Description.} Given an interaction graph $\mathcal{G}_u$ that captures the historical interactions between users and items, as well as an external knowledge graph $\mathcal{G}_k$ that depicts relationships between various entities, the key task we aim to address involves learning an effective function that can predict the relative likelihood of a user adopting a specific item with which they have not previously interacted. More formally, our goal is to train a model that can accurately quantify the probability that a given user $u$ will positively interact with a candidate item $i$, given the user's interaction history in $\mathcal{G}_u$ and the wealth of relational information encoding various dependencies and correlations in $\mathcal{G}_k$. By successfully encoding both sources of data, the model can score all unseen user-item pairs to generate personalized rankings over the unseen items per user for accurate recommendations. The primary challenge lies in effectively integrating the connectivity structure and semantic information within the auxiliary knowledge graph $\mathcal{G}_k$ at a fine-grained level to enhance the representations learned for users and items solely from $\mathcal{G}_u$, thereby boosting predictive performance. Overcoming this challenge will allow us to develop improved recommender systems.

\section{METHODOLOGY}
\begin{figure*}[tbp]
\centering
  \includegraphics[scale=0.71]{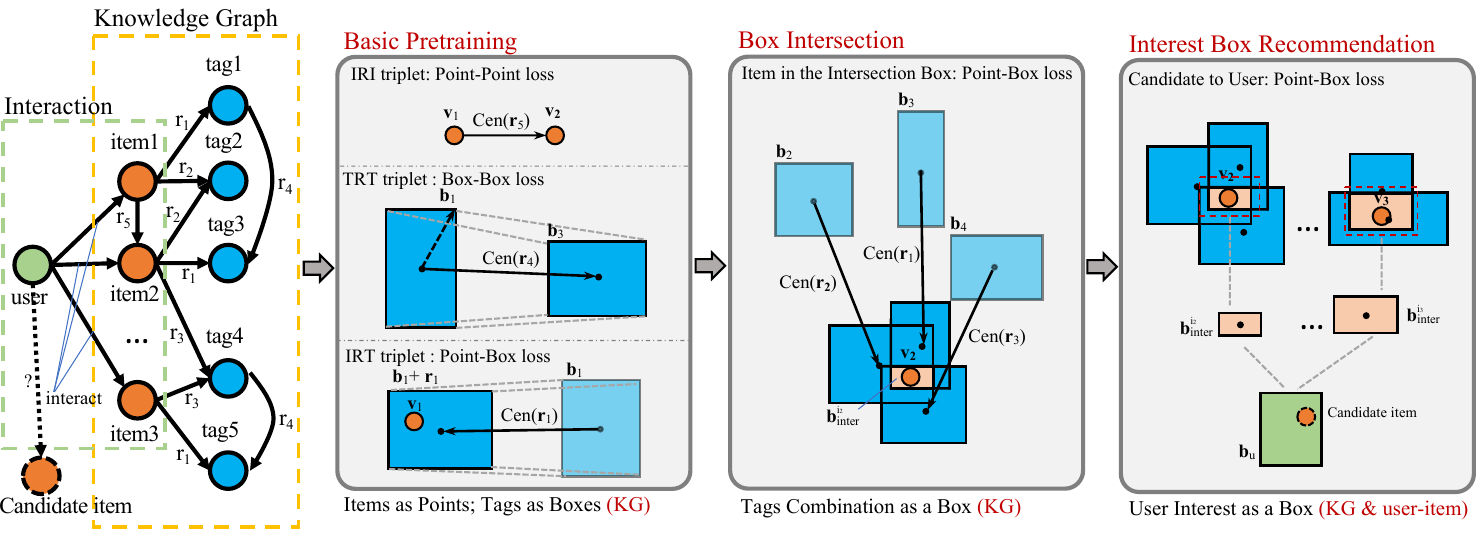}
  \caption{Framework of the proposed model {\model}: The recommendation task is completed through three training steps (not all required). The initial two steps focus solely on the KG data to obtain suitable representations for items, tags, and relations. In the third step, the objective is to leverage the user's interest box to compute the matching score, which serves as the recommendation result.}
  \label{fig:framework}
\end{figure*}
In this section, we introduce the details of the proposed model, and Figure~\ref{fig:framework} illustrates the architecture of the whole model. Firstly, we describe the representation format of items, tags, and relations in the KG. To train these different forms of embeddings and use them to get the recommendation results, we design three training steps to optimize the embeddings: (1) Basic pretraining, which aims to obtain suitable representations for each item, tag, and relation in the KG. To achieve this goal, three distinct distance functions are employed corresponding to the different triplet types; (2) Box intersection step, which seeks to embed each item point within the intersection region of its associated tag boxes. This meets our intuitive requirements when defining this representation; and (3) Interest box recommendation, which aspires to obtain the user interest box with the related tags' box embeddings, and employ the interest box embedding to get the predicted scores for all items the user has not interacted with, so that we can rank the items for recommendation results.

\subsection{Item Point and Relation, Tag Box}
To efficiently connect a set of items with a related concept in the embedding space, we first define point and box embeddings (i.e., axis-aligned hyper-rectangles) for the items and tags in the knowledge graph, respectively. Formally, in a vector space with dimension $d$, let each item be a point in the space, which could be represented as a vector $\mathbf{v} \in \mathbb{R}^{d}$. 
A tag in KG could be regarded as a box so that it could encompass multiple points in space that represent items. We define a box embedding as $\mathbf{b} = (\text{Cen}(\mathbf{b}), \text{Off}(\mathbf{b})) \in \mathbb{R}^{2d}$, where $\text{Cen}(\mathbf{b})$ is the center point of the box, and $\text{Off}(\mathbf{b})$ is the offset embedding of the box, which indicates the size of the box. Through the calculation of two vectors, we can get the range of the box embedding on each dimension, so the box embedding $\mathbf{b}$ specifies the box range in the vector space as follows:

\begin{equation}
\label{box_range}
\text{Range}(\mathbf{b}) = { \text{Cen}(\mathbf{b}) - \sigma(\text{Off}(\mathbf{b})) \preceq \mathbf{v} \preceq \text{Cen}(\mathbf{b}) + \sigma(\text{Off}(\mathbf{b})) }
\end{equation}

where $\sigma$ is an activation function (e.g., \texttt{ReLu}) to make the $\text{Off}(\mathbf{b})$ be positive, and $\preceq$ is element-wise inequality. According to the definition, a box could contain all the items with their point embeddings inside the box range, building the connectivity of a concept and a set of items, which is one of the benefits of regarding a tag as a box embedding compared to a single point.

\subsection{Basic Pretraining Step}

Building upon the definitions of box embedding and point embedding, we can map each item and tag in the KG as a point or a box in the space. As previously mentioned, the same tag with different relations may represent distinct concepts(like \textit{(is directed by, James Francis Cameron)} and \textit{(is written by, James Francis Cameron)})). To obtain appropriate concept embeddings, it is crucial to capture the effect of the relation on tags within the vector space. To achieve this, we have to define the representation of a relation as a box $\mathbf{b}_{r} = (\text{Cen}(\mathbf{b}_{r}), \text{Off}(\mathbf{b}_{r})) \in \mathbb{R}^{2d}$, which is identical to a tag, and the relation box embedding will be used to modify the tag box embedding. When examining its association with the tag box, $\text{Cen}(\mathbf{b}_{r})$ denotes the projection operator for the tag box's central point, while $\text{Off}(\mathbf{b}_{r})$ signifies the resizing of the tag box's dimensions. Importantly, when evaluating the relation's impact on items, we exclusively employ $\text{Off}(\mathbf{b}_{r})$, as a point lacks the notion of boundaries.

\begin{figure}[tbp]
\centering
  \includegraphics[scale=1.01]{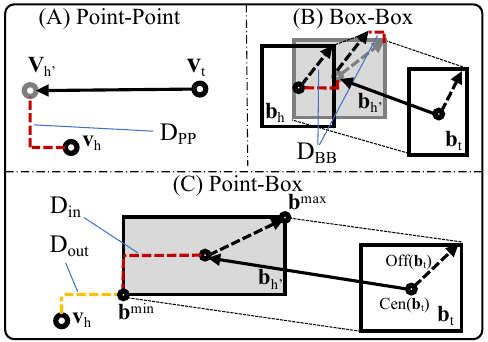}
  \caption{The geometric intuition of the different distances in 2-dimensional space. (A) The point-to-point distance for \textit{IRI} triplets. (B) The box-to-box distance for \textit{TRT} triplets. (C) The inside and outside distance for \textit{IRT} triplets.}
  \label{fig:distance}
\end{figure}

After the embedding is initialized randomly, it is necessary to get a representation that satisfies the triplets in KG.
In the basic pretraining step, the main target is locating the item point inside each box of its related concept box, and the key idea is like the translation-based KG embedding models using relation as the projector. As we mentioned before, there are three kinds of triplets in the KG, which are \textit{IRI} triplet, \textit{TRT} triplet, and \textit{IRT} triplet. Since each type of triplet contains a different head and tail entity embedding form, we design different ways of calculating the distance to get the final loss according to the type of the triplets, as Figure~\ref{fig:distance} illustrates the different calculation processes.
For \textit{IRI} triplets, we can simply use the point-to-point distance between the head and tail entities, as they are both represented as points. For \textit{TRT} triplets, since both the head and tail entities are represented as boxes, we can use the box-to-box distance to calculate their similarity. Finally, for the \textit{IRT} triplets, the head entity is a point, while the tail entity is a box, so in this case, we could use the point-to-box distance to measure their similarity.
By calculating these distances, we can derive a loss function for each type of triplet. The goal of the basic pretraining step is to minimize the total loss, which will ensure that the item points are located inside the corresponding boxes and properly represent the facts in the KG.

\textbf{IRI triplets.} For this kind of triplet, the head entity and tail entity are both represented as points, and the object is to make the tail point close enough to the head point after projection with the relation. 
Let's take the example \textit{(Avatar2, is the sequel of, Avatar)} in Section 2 as a demonstration. For this \textit{IRI} triplet, the objective is to ensure that the object representation of \textit{Avatar} closely aligns with the point representation of \textit{Avatar2} after mapping by relation \textit{is sequel of}.
Specifically, for a triplet $(h, r, t)$ ($h$ and $t$ are both items), the functional mapping with relation $r$ is an element-wise addition from $t$ to the target item $h$ and gets the predicted embedding as:
\begin{equation}
    \label{IRI}
    \mathbf{v}_{h}^{\prime} = \mathbf{v}_{t} + \text{Cen}(\mathbf{b}_{r})
\end{equation}
Note that the relation is utilized as a projector to the tail entity $t$ for compliance with the other two kinds of triplets, which will be introduced later. 
Essentially, there is no difference with projection on head entity $h$.
The distance function is defined as follows:
\begin{equation}
    \label{distance_IRI}
    \mathbf{D}_{PP}(\mathbf{v}_{h}, \mathbf{v}_{h}^{\prime}) =  \left\Vert \mathbf{v}_{h} - \mathbf{v}_{h}^{\prime} \right\Vert_{1}
\end{equation}
where $\left\Vert \mathbf{x} \right\Vert_{1}$ means the L1 norm function and $\mathbf{D}_{PP}$ means the point-to-point distance. 

In summary, for IRI triplets, we compute the distance between two points in the embedding space, which ensures that the learned embeddings accurately represent the connections between different items in the KG.

\textbf{TRT triplets.} The \textit{TRT} triplets reflect the connection between two tags. For such a triplet, its head and tail tags are both represented by a box, when using the relation for projection, it is necessary to consider not only the tag box center point translation but also modifying the tag box offset size. For example, for the triplet \textit{(James Francis Cameron, is a citizen of, America)}, the objective is to ensure that the object box, mapped from \textit{America} with relation, close to the box of \textit{James Francis Cameron}. We utilize the center embedding and offset embedding of relation $\text{Cen}(\mathbf{b}_{r}), \text{Off}(\mathbf{b}_{r})$ to achieve this goal as follows:
\begin{equation}
    \label{TRT_cen}
    \text{Cen}(\mathbf{b}_{h}^{\prime}) = \text{Cen}(\mathbf{b}_{t}) + \text{Cen}(\mathbf{b}_{r})
\end{equation}
\begin{equation}
    \label{TRT_off}
    \text{Off}(\mathbf{b}_{h}^{\prime}) = \sigma(\text{Off}(\mathbf{b}_{t})) + \text{Off}(\mathbf{b}_{r})
\end{equation}
where the $\sigma$ is the \texttt{ReLu} function in our paper since the box offset embedding of a tag $\text{Off}(\mathbf{b}_{t})$ should be positive, while the element of $\text{Off}(\mathbf{b}_{r})$ could be negative so it could be used to narrow $\sigma(\text{Off}(\mathbf{b}_{t}))$. The first operation~\ref{TRT_cen} means changing the position of the box center point, while the second operation~\ref{TRT_off} means adjusting the size of the box.

The distance function of the TRT triplet should consider the similarity of the center and offset embedding at the same time, which is defined as follows:
\begin{equation}
     \label{distance_TRT}
\begin{split}
    \mathbf{D}_{BB}(&\mathbf{b}_{h}, \mathbf{b}_{h}^{\prime}) = \left\Vert \text{Cen}(\mathbf{b}_{h}) - \text{Cen}(\mathbf{b}_{h}^{\prime}) \right\Vert_{1} + \\
    &\left\Vert \sigma(\text{Off}(\mathbf{b}_{h})) - \sigma(\text{Off}(\mathbf{b}_{h}^{\prime})) \right\Vert_{1}
\end{split}
\end{equation}
where $\mathbf{D}_{BB}$ measures the distance between two box embeddings from the perspectives of center position deviation and size deviation.

In summary, for TRT triplets, we account for both the center and offset embeddings in calculating the similarity between boxes. This ensures that the learned embeddings accurately represent the connections between tags in the KG and allows for the consideration of the center and size of the boxes simultaneously.

\textbf{IRT triplets.} The \textit{IRT} triplet contains an item and a tag, and the target is to locate the item point into the range of the concept box, which is generated by projecting the tag box using the relation box embedding. For a triplet $(h, r, t)$ ($h$ is an item and $t$ is a tag), since even for the same tag like \textit{James Francis Cameron}, different relation-tag pairs like \textit{(is written by, James Francis Cameron)} and \textit{(is directed by, James Francis Cameron)} are different concepts, the center embedding and offset embedding of relation $\text{Cen}(\mathbf{b}_{r}), \text{Off}(\mathbf{b}_{r})$ are used to translate the tag box center $\text{Cen}(\mathbf{b}_{t})$ and modify the offset $\text{Off}(\mathbf{b}_{t})$, respectively, just as the operation for \textit{TRT} triplets with Equation~\ref{TRT_cen} and \ref{TRT_off}.
For example, for the \textit{IRT} triplet \textit{(Avatar, is directed by, James Francis Cameron)}, the objective is to ensure that the concept box, mapped from \textit{James Francis Cameron} with relation \textit{is directed by$^{-1}$}, could contain the point of \textit{Avatar}.

In order to measure whether the vector after projection is appropriate, we should define two types of distance between a point and a box, which are the outside distance and the inside distance:
\begin{equation}
    \label{distance_IRT}
    \mathbf{D}_{PB}(\mathbf{v}_{h}, \mathbf{b}_{h}^{\prime}) =  \mathbf{D}_{out}(\mathbf{v}_{h}, \mathbf{b}_{h}^{\prime}) + \mathbf{D}_{in}(\mathbf{v}_{h}, \mathbf{b}_{h}^{\prime})
\end{equation}
where $\mathbf{D}_{PB}$ denotes point-to-box distance, while $\mathbf{D}_{out}(\mathbf{v}, \mathbf{b})$ and $\mathbf{D}_{in}(\mathbf{v}, \mathbf{b})$ is the outside distance and the inside distance, respectively. Note that the box is defined as an axis-aligned rectangle, so the two types of distance are calculated according to each dimension in the vector space as follows, which are defined as follows:
\begin{equation}
    \label{distance_out}
    \mathbf{D}_{out}(\mathbf{v}, \mathbf{b}) = \left\Vert \texttt{Max}(\mathbf{v} - \mathbf{b}^{max}, \mathbf{0}) \right\Vert_{1} + \left\Vert \texttt{Min}(\mathbf{b}^{min} - \mathbf{v}, \mathbf{0}) \right\Vert_{1}
\end{equation}
\begin{equation}
    \label{distance_in}
    \mathbf{D}_{in}(\mathbf{v}, \mathbf{b}) = \left\Vert \text{Cen}({\mathbf{b}}) - \texttt{Min}(\mathbf{b}^{max}, \texttt{Max}(\mathbf{b}^{min}, \mathbf{v})) \right\Vert_{1} 
\end{equation}
where \texttt{Max} and \texttt{Min} are element-wise functions, which means selecting the larger or smaller element for each dimension. $\mathbf{b}^{max}$ and $\mathbf{b}^{min}$ are the points with the largest or smallest value in each dimension, which are calculated as follows:
\begin{equation}
    \label{max_point}
    \mathbf{b}^{max} = \text{Cen}({\mathbf{b}}) + \sigma(\text{Off}({\mathbf{b}}))
\end{equation}
\begin{equation}
    \label{min_point}
    \mathbf{b}^{min} = \text{Cen}({\mathbf{b}}) - \sigma(\text{Off}({\mathbf{b}}))
\end{equation}
According to the above definition, it can be seen that the point-to-box distance is used to measure the distance between a point and a box embedding calculating the distance between that point and the center point of the box and the closest point to the box border from each dimension.

In summary, for IRT triplets, $\mathbf{D}_{out}(\mathbf{v}_{h}, \mathbf{b}_{t}^{\prime})$ quantifies the proximity between point $\mathbf{v}_{h}$ and the nearest point within the box boundary, while $\mathbf{D}_{in}(\mathbf{v}_{h}, \mathbf{b}_{t}^{\prime})$ assesses the distance from the point to the box's center (provided the point resides inside the box along that dimension) or the offset magnitude (in the case that the point lies outside the box along that dimension).

\textbf{Training and Optimization.} In this step, we construct a loss function that considers positive and negative samples with different sample weights to optimize the model. Specifically, for an \textit{IRI} or \textit{TRT} triplet to be predicted $(?, r, t)$ in the KG, $n$ negative head entities will be randomly selected, which are not connected with $t$ via relation $r$. For an \textit{IRT} triplet $(i, r, t)$, there are two approaches to generate a negative sample: replacing the item $i$ or tag $t$. With the true triplets being positive samples, the goal of optimization is to minimize the distance of the positive samples and maximize the distance of negative samples. Therefore, the loss function is defined as:

\begin{equation}
    \label{loss}
    \mathcal{L} = - w * (\text{log} \theta(\gamma - \text{D}^{pos}) - \frac{1}{n}\sum_{i=1}^{n}\text{log} \theta(\gamma - \text{D}^{neg}))
\end{equation}

where $\theta$ is the \texttt{sigmoid} function, $\gamma$ is a fixed scalar margin, $n$ is the size of the negative sample, and $w$ is the sample weight. The more correct answers that exist, the smaller $w$ is, and it is used to balance the importance of each triplet in a training batch. $\text{D}^{pos}$ and $\text{D}^{neg}$ denote the distance of positive and negative triplets, which are calculated with Equation~\ref{distance_IRI}, \ref{distance_TRT}, or \ref{distance_IRT} according to the triplet type. For each step, one type of triplet will be sampled, and the possibility is decided by its proportion in all types.

\subsection{Box intersection}

Beyond merely necessitating alignment between the item point and the concept box with respect to the triplets in the KG, it is also imperative for each item to reside within the intersection region formed by the amalgamation of its associated concepts box. Indeed, we posit that this composite concept encapsulates accurate and complex attributes of an item. While the Basic Pretraining step positions the item within each concept box, this intersection criterion is only implicitly fulfilled. Consequently, we carry out a secondary training phase with the objective of obtaining the intersection of a collection of boxes and situating the item point within the resultant intersecting box.

Let $i$ represent the item, with its point embedding denoted as $\mathbf{v}$. Additionally, we obtain the tags ${t_{1}, t_{2},...,t_{n}}$ and relations ${r_{1}, r_{2},...,r_{n}}$ associated with item $i$ based on the knowledge graph. The sets of tag and relation embeddings are represented as ${\mathbf{b}_{t_{1}}, \mathbf{b}_{t_{2}},...,\mathbf{b}_{t_{n}}}$ and ${\mathbf{b}_{r_{1}}, \mathbf{b}_{r_{2}},...,\mathbf{b}_{r_{n}}}$. Utilizing the projection operation in accordance with Equation~\ref{TRT_cen} and \ref{TRT_off} applied to the tag and relation embeddings, we obtain a final set of concept boxes ${\mathbf{b}_{1}, \mathbf{b}_{2},...,\mathbf{b}_{n}}$. To ascertain the intersecting box using the embeddings, there are two approaches: one employing an attention neural network and the other a purely mathematical method. These are referred to as Attention Network Intersection and Max-Min Intersection, respectively.

\textbf{Attention Network Intersection.} The fundamental concept involves utilizing a neural network to determine the attention of each box center, as different boxes may have varying influences on the intersection region. The center embedding of $\mathbf{b}{inter}$ is defined as follows:
\begin{equation}
    \label{center_intersection}
    \text{Cen}(\mathbf{b}_{inter}) = \sum_{i=1}^{n} \mathbf{a}_{i} \circ \text{Cen}(\mathbf{b}_{i})
\end{equation}
where $\mathbf{a}_{i} \in \mathbb{R}^{d}$ represents the attention embedding and $\circ$ denotes element-wise multiplication. The core principle of Equation~\ref{center_intersection} is to allocate an attention value to each box for every dimension. $a_{i}$ is generated using a \texttt{MLP} function:
\begin{equation}
    \label{a_generation}
    \mathbf{a}_{i} = \frac{\text{exp}(\text{MLP}(\text{Cen}(\mathbf{b}_{i}))}{\sum_{j=1}^{n}\text{exp}(\text{MLP}(\text{Cen}(\mathbf{b}_{j}))}
\end{equation}

For the offset embedding of the intersection box, it should be smaller than any offset in ${\mathbf{b}_{1}, \mathbf{b}_{2},...,\mathbf{b}_{n}}$ in each dimension. Thus, we select the minimal element and generate a shrinking scale for each dimension with the offset embedding set ${\text{Off}(\mathbf{b}_{1}), \text{Off}(\mathbf{b}_{2}),...,\text{Off}(\mathbf{b}_{n}) }$:
\begin{equation}
    \label{offset_intersection}
    \text{Off}(\mathbf{b}_{inter}) = \text{Min}(\sigma(\text{Off}(\mathbf{b}_{1})), ...,\sigma(\text{Off}(\mathbf{b}_{n}))) \circ \mathbf{g}
\end{equation}

where $\texttt{Min}$ is an element-wise minimal function, and $\mathbf{g} \in \mathbb{R}^{d}$ represents the shrinking embedding, generated as follows:
\begin{equation}
    \label{g_generation}
    \mathbf{g} = \theta(\text{MLP}(\frac{1}{n}\sum_{i=1}^{n}\text{MLP}(\text{Off}(\mathbf{b}_{i}))))
\end{equation}

\textbf{Max-Min Intersection.} Since we are generating the intersection of multiple box embeddings, a more intuitive approach is to directly determine the intersection region across all the boxes in each dimension by selecting the maximal or minimal value from ${\mathbf{b}_{1}, \mathbf{b}_{2},...,\mathbf{b}_{n}}$. Specifically, for the intersection box, we can obtain the max point $\mathbf{b}_{inter}^{max}$, defined in Equation~\ref{max_point}, by selecting the minimal value from each box's max point in each dimension. Similarly, $\mathbf{b}_{inter}^{min}$ should choose the maximal value from the min points:
\begin{equation}
    \label{max_intersection}
    \mathbf{b}_{inter}^{max} = \texttt{Min}(\sigma(\mathbf{b}_{1}^{max}), \sigma(\mathbf{b}_{2}^{max}),...,\sigma(\mathbf{b}_{n}^{max}) )
\end{equation}
\begin{equation}
    \label{min_intersection}
    \mathbf{b}_{inter}^{min} = \texttt{Max}(\sigma(\mathbf{b}_{1}^{min}), \sigma(\mathbf{b}_{2}^{min}),...,\sigma(\mathbf{b}_{n}^{min}) )
\end{equation}
where the $\texttt{Max}$ and $\texttt{Min}$ are element-wise function just like Equation~\ref{distance_out}.
According to Equation~\ref{max_point} and \ref{min_point}, with $\mathbf{b}_{inter}^{max}$ and $\mathbf{b}_{inter}^{min}$, we could calculate the $\text{Cen}(\mathbf{b}_{inter})$ and $\text{off}(\mathbf{b}_{inter})$ as follows:
\begin{equation}
    \label{center_intersection1}
    \text{Cen}(\mathbf{b}_{inter}) = (\mathbf{b}_{inter}^{max} + \mathbf{b}_{inter}^{min}) / 2
\end{equation}
\begin{equation}
    \label{offset_intersection1}
    \text{Off}(\mathbf{b}_{inter}) = \sigma(\mathbf{b}_{inter}^{max} - \mathbf{b}_{inter}^{min}) / 2
\end{equation}
 By the above method, we can obtain the embedding of the intersection region of multiple boxes using simple mathematical operations without neural networks.

\textbf{Training and Optimization.} In this step, given an item and its associated relation-tag pair, the intersection box embedding $\mathbf{b}_{inter} = (\text{Cen}(\mathbf{b}_{inter}), \text{Off}(\mathbf{b}_{inter}))$ can be obtained with the two methods mentioned above. Using the item embedding $\mathbf{v}$ and the intersection box embedding $\mathbf{b}_{inter} = (\text{Cen}(\mathbf{b}_{inter}), \text{Off}(\mathbf{b}_{inter}))$, the optimization goal is to position the item point within the box. The distance function is similar to Equation~\ref{distance_IRT}, which we employed for \textit{IRT} triplets in the basic pretraining step.
Negative samples are generated by replacing the item with other randomly selected items that are not related to these relation-tag pairs, and the loss function is identical to Equation~\ref{loss}. The sample weight $w$ in the loss function is defined as $w = 1 / (n + 1)$, where $n$ represents the number of concepts related to the item.

\subsection{Interest Box Recommendation}
The preceding two steps were trained only on the KG, with the objective of obtaining suitable representations for items, tags, and relations. Building on this foundation, we will utilize the pre-trained embedding for making predictions in the recommendation task. As previously discussed, a user's interests can be intricate and precise, making them challenging to represent with manually defined concepts like \textit{a dress}. Instead, they may encompass more complex concepts, such as \textit{a red prom dress} (encompassing three aspects: red, prom, and dress), or even more intricate combinations of broader concepts. This amalgamation resembles the intersection of boxes. For example, a user could interact with an item $i$ and an item $j$, and the combination concept of $i$ and $j$ is \textit{red prom dress} and \textit{prom highheels}, respectively. We could infer that the interest of this user may be interested in \textit{prom outfit}.

In this step, each item's intersection box region is employed to generate the user's interest box. Since the influence of the same item may differ among users, a user-bias box intersection is conducted, which works in conjunction with the intersection results from the second training step. Ultimately, the user's interests are represented by a box embedding that integrates all items from the interaction history and is used to rank candidate items.

\textbf{User-bias Intersection.} For a user $u$ assigned an embedding $\mathbf{u} \in \mathbb{R}^{d}$, the items in the intersection history form a set ${i_{1}, i_{2},..., i_{m}}$. In the second step, the intersection box of these items, denoted as $\mathbf{b}_{interI}$ (\textit{interI} represents intersection considering only the item), is generated without taking the user into account. To incorporate the user, we concatenate the user embedding $\mathbf{u}$ with each concept box center embedding $\text{Cen}(\mathbf{b})$ and offset embedding $\text{Off}(\mathbf{b})$, and use the combined embeddings to obtain the intersection box $\mathbf{b}_{interU}$ (\textit{interU} represents intersection considering the user) as follows:

\begin{equation}
    \label{center_intersection2_1}
    \text{Cen}(\mathbf{b}_{interU}) = \sum_{i=1}^{n} \mathbf{c}_{i} \circ \text{Cen}(\mathbf{b}_{i}),
\end{equation}
\begin{equation}
    \label{center_intersection2_2}
    \text{Off}(\mathbf{b}_{interU}) = \sum_{i=1}^{n} \mathbf{d}_{i} \circ \text{Off}(\mathbf{b}_{i})
\end{equation}

$\mathbf{c}_{i}, \mathbf{d}_{i} \in \mathbb{R}^{d}$ are the attention embeddings generated as:

\begin{equation}
    \label{c_generation_1}
    \mathbf{c}_{i} = \frac{\text{exp}(\text{MLP}(\text{Cen}(\mathbf{b}_{i}), \mathbf{u})}{\sum_{j=1}^{m}\text{exp}(\text{MLP}(\text{Cen}(\mathbf{b}_{j}), \mathbf{u})}
\end{equation}
\begin{equation}
    \label{c_generation_2}
    \mathbf{d}_{i} = \frac{\text{exp}(\text{MLP}(\text{Off}(\mathbf{b}_{i}), \mathbf{u})}{\sum_{j=1}^{m}\text{exp}(\text{MLP}(\text{Off}(\mathbf{b}_{j}), \mathbf{u})}
\end{equation}

where \text{MLP}($\cdot$): $\mathbb{R}^{2d} \rightarrow \mathbb{R}^{d}$ transforms the $2d$-dimensional vector into a $d$-dimensional vector, which is distinct from the MLP in Equation~\ref{a_generation} and \ref{g_generation}.

\textbf{Interest Box.} With the two intersection boxes for each item, $\mathbf{b}_{interU}$ and $\mathbf{b}_{interI}$, the user's interest can be represented by the intersection box embeddings of their interacted items: ${ \mathbf{b}_{interU_1}, \mathbf{b}_{interU_2},...,\mathbf{b}_{interU_m}}$ and ${ \mathbf{b}_{interI_1}, \mathbf{b}_{interI_2},...,\mathbf{b}_{interI_m}}$. For each item, its final box embedding is the average of $\mathbf{b}_{interU}$ and $\mathbf{b}_{interI}$:

\begin{equation}
    \label{item_box_center}
    \text{Cen}(\mathbf{b}_{inter}) = (\text{Cen}(\mathbf{b}_{interI}) + \text{Cen}(\mathbf{b}_{interU})) / 2
\end{equation}
\begin{equation}
    \label{item_box_offset}
    \text{Off}(\mathbf{b}_{inter}) = (\text{Off}(\mathbf{b}_{interI}) + \text{Off}(\mathbf{b}_{interU})) / 2
\end{equation}

The user $u$'s interest considers all of the item boxes. With the item box embedding, we could form the user interest box embedding as follows:

\begin{equation}
    \label{user_box_1}
    \text{Cen}(\mathbf{b}_{u}) = \frac{1}{m}\sum_{k=1}^{m}\text{Cen}(\mathbf{b}_{inter\_k})
\end{equation}
\begin{equation}
    \label{user_box_2}
    \text{Off}(\mathbf{b}_{u}) = \frac{1}{m}\sum_{k=1}^{m}\text{Off}(\mathbf{b}_{inter\_k})
\end{equation}

This user interest box $\mathbf{b}_{u}$ is used to rank all items that have never been interacted with by the user according to the point-to-box distance.

\textbf{Training and Optimization.} For the recommendation task, the model generates the interest box for each user as described above. The optimization goal is to locate the interacted items' point embedding in the user's interest box while keeping the items that have not been engaged with the user outside of the box. Therefore, the point-to-box distance function $\mathbf{D}_{PB}(\mathbf{v}, \mathbf{b}_{u})$ is adopted in this step. The negative samples are randomly selected from the items that are not in the interacted set ${i_{1}, i_{2},...,i_{m}}$. The loss function is the same as Equation~\ref{loss}, and the sample weight is defined as $w = 1/(m + \alpha)$, where $m$ is the size of the interacted item set and $\alpha$ is a scale to balance the weight.

\subsection{Model Prediction}
After the three-stage training process, we obtain a point representation vector $\mathbf{v}_i$ for each item $i$ and an interest box representation $\mathbf{b}_u$ for each user $u$ encoded in the same embedding space. Now given a specific user-item pair $(u, i)$ for recommendation, we can calculate an interest matching score between them using the proposed point-to-box distance function as follows:

\begin{equation}
    \label{score}
    \text{Score}(\mathbf{v}_{i}, \mathbf{b}_{u}) = \gamma - \text{D}_{PB}(\mathbf{v}_{i}, \mathbf{b}_{u})
\end{equation}

Where $\gamma$ is a predefined scale hyperparameter used for calibrating the box representations in Equation~\ref{loss}. Intuitively, this scoring function quantifies the geometric proximity between an item's point representation $\mathbf{v}_i$ and the user's personal interest box $\mathbf{b}u$ in the embedding space. A smaller point-to-box distance $\text{D}{PB}$ indicates $\mathbf{v}_i$ lies closer to or within $\mathbf{b}_u$, implying a higher interest match. By subtracting the distance from a large constant $\gamma$, we convert it to a normalized matching score. For recommendation generation, we can simply rank all candidate items by their Score($\mathbf{v}_i, \mathbf{b}_u$) in descending order. Items with higher scores will be ranked at the top as they exhibit closer alignment with the user's interests. In this way, our box-based representation learning approach allows for generating personalized recommendations for each user by performing a geometric search in the embedding space to retrieve items closest to their interest boxes.

\section{EXPERIMENTS}
In this segment, we embark on experiments utilizing four real-world datasets spanning diverse domains to thoroughly assess the efficacy of our proposed model. The empirical findings are geared towards addressing the following pivotal research inquiries:

\begin{itemize}
\item \textbf{RQ1}: How does our model compare against state-of-the-art recommendation baselines? This investigates the overall performance of integrating knowledge graph information to enhance recommendations.

\item \textbf{RQ2}: What are the impacts of different components in our framework, including multi-stage training methodology, diverse data types, intersection mechanisms, etc? This provides an in-depth analysis of which design factors contribute the most to the performance gains.

\item \textbf{RQ3}: Can we qualitatively validate if the box constraints successfully make items sharing common conceptual attributes have closer distributed representations as expected? This examines whether the knowledge graph integration achieves the goal of clustering semantically similar items in the embedding space.
\end{itemize}

Through comprehensive quantitative comparisons and carefully designed ablation studies towards answering the above key questions, we aim to validate the effectiveness of the knowledge graph-enhanced recommendation model proposed in this paper and provide insights into the best practices of integrating structured knowledge into recommender systems.

\subsection{Experimental Settings}

\subsubsection{Datasets Description.}
We conduct our experiments on four diverse real-world datasets, extensively employed in prior research, to assess the comprehensive performance of {\model}: 

\textbf{Last-FM} dataset is a valuable resource in recommendation system research, offering comprehensive user listening histories along with metadata on artists, albums, and songs.

\textbf{Yelp2018} for business venue recommendation. Where local businesses like restaurants and bars are viewed as items.

\textbf{Alibaba-iFashion} Alibaba-iFashion dataset is a fashion clothing dataset collected from the Alibaba online shopping system.

\textbf{Amazon-Book} dataset is a collection of information on books sold on Amazon, including details such as title, author, price, ratings, reviews, and publication date.

As some cutting-edge models are based on GNN and require multi-hop neighbors of items in the KG, the datasets include two-hop neighbors to construct the knowledge graph. The overall statistics, encompassing user-item interaction history and the knowledge graph, are summarized in Table~\ref{table:datasets}, and note that these datasets do not distinguish the interaction type. In accordance with our previous classification of KG triplet types in Section 2, we separately count the number of different triplet types and their proportions, which will prove beneficial for our experimental analysis.

\begin{table}[ht]
\centering
\caption{Statistics of the datasets used in our experiments. We report the basic statistics about the datasets, and notably, we also report the proportion of different types of triplets present in the KG.}
\label{table:datasets}
\begin{tabular}{lrrrr} 
\toprule
\textbf{Stas.}           & \multicolumn{1}{c}{\begin{tabular}[c]{@{}c@{}}\textbf{Last}\\\textbf{-FM}\end{tabular}} & \multicolumn{1}{c}{\begin{tabular}[c]{@{}c@{}}\textbf{Yelp}\\\textbf{2018}\end{tabular}} & \multicolumn{1}{c}{\begin{tabular}[c]{@{}c@{}}\textbf{Alibaba-}\\\textbf{iFashion}\end{tabular}} & \multicolumn{1}{c}{\begin{tabular}[c]{@{}c@{}}\textbf{Amazon}\\\textbf{-book}\end{tabular}}  \\ 
\midrule
                         & \multicolumn{4}{c}{\textbf{User-Item Interaction}}                                                                                                                                                                                                                                                                                                                                   \\ 
\hline
\textbf{\#Users}         & 23,566                                                                                  & 45,919                                                                                   & 114,737                                                                                          & 70,679                                                                                       \\
\textbf{\#Items}         & 48,123                                                                                  & 45,538                                                                                   & 30,040                                                                                           & 24,915                                                                                       \\
\textbf{\#Intersections} & 3,034,796                                                                               & 1,185,068                                                                                & 1,781,093                                                                                        & 847,733                                                                                      \\ 
\hline
                         & \multicolumn{4}{c}{\textbf{Knowledge Graph}}                                                                                                                                                                                                                                                                                                                                         \\ 
\hline
\textbf{\#Items}         & 48,123                                                                                  & 45,538                                                                                   & 30,040                                                                                           & 24,915                                                                                       \\
\textbf{\#Tags}          & 58,266                                                                                  & 90,961                                                                                   & 59,156                                                                                           & 88,572                                                                                       \\
\textbf{\#Relations}     & 9                                                                                       & 42                                                                                       & 51                                                                                               & 39                                                                                           \\
\textbf{\#IRI Triplets}  & 3,284                                                                                   & 0                                                                                        & 0                                                                                                & 2,985                                                                                        \\
\textbf{\#TRT Triplets}  & 113,546                                                                                 & 984,101                                                                                  & 173,690                                                                                          & 1,868,245                                                                                    \\
\textbf{\#IRT Triplets}  & 347,737                                                                                 & 869,603                                                                                  & 105,465                                                                                          & 686,516                                                                                      \\
\textbf{IRI (\%)}        & 0.71\%                                                                                  & 0.00\%                                                                                   & 0.00\%                                                                                           & 0.12\%                                                                                       \\
\textbf{TRT (\%)}        & 24.44\%                                                                                 & 53.09\%                                                                                  & 62.22\%                                                                                          & 73.04\%                                                                                      \\
\textbf{IRT (\%)}        & 74.85\%                                                                                 & 46.91\%                                                                                  & 37.78\%                                                                                          & 26.84\%                                                                                      \\
\bottomrule
\end{tabular}
\end{table}

\subsubsection{Evaluation Metrics.}
During the evaluation phase, we adopt the all-ranking strategy~\cite{SM, KGIN} to ensure a fair comparison. Specifically, for each target user, all items not yet interacted with are considered negative, while their interacted items in the test set are deemed positive for inferring user preferences. All these items are ranked according to the matching scores produced by the model. To evaluate performance, we employ the established protocols~\cite{SM}: recall@$K$ and ndcg@$K$, with $K=20$ as the default value. The experiment results report the average metrics for all users in the test set.

\begin{table*}[t]
\setlength\tabcolsep{3pt} 
\centering
\caption{Overall Results of Our Model on Different Datasets and Comparison with Baseline Models. This table reports the NDCG and Recall metrics of different methods on various datasets. The best results are highlighted in bold and the second best results are underlined. The numbers in brackets indicate the relative improvements of our proposed method over the baseline models. The results reported for the baseline models are taken from prior published work.}
\label{table:results}
\resizebox{\textwidth}{!}{ 
\begin{tabular}{ccccccccc} 
\toprule
                    & \multicolumn{2}{c}{\textbf{Last-FM}}                  & \multicolumn{2}{c}{\textbf{Yelp2018}}               & \multicolumn{2}{c}{\textbf{Alibaba-iFashion}}       & \multicolumn{2}{c}{\textbf{Amazon-book}}             \\ 
\hline
                    & \textbf{recall}           & \textbf{ndcg}             & \textbf{recall}          & \textbf{ndcg}            & \textbf{recall}          & \textbf{ndcg}            & \textbf{recall}          & \textbf{ndcg}             \\ 
\midrule
\textbf{MF}         & 0.0724 (57.46\%)         & 0.0617 (83.79\%)         & 0.0627 (28.55\%)        & 0.0413 (21.31\%)        & 0.1095~ (21.92\%)       & 0.067 (26.57\%)         & 0.1300 (34.77\%)        & 0.0678 (41.15\%)         \\
\textbf{CKE}        & 0.0732 (55.74\%)         & 0.0630 (80.00\%)         & 0.0653 (23.43\%)        & 0.0423 (18.44\%)        & 0.1103 (21.03\%)        & 0.0676 (25.44\%)        & 0.1342 (30.55\%)        & 0.0698 (37.11\%)         \\
\textbf{UGRec}      & 0.0730 (56.16\%)         & 0.0624 (81.73\%)         & 0.0651 (23.81\%)        & 0.0419 (19.57\%)        & 0.1006 (32.70\%)        & 0.0621 (36.55\%)        & -                        & -                         \\ 
\textbf{H-know} & 0.0948 (20.25\%)         & 0.0812 (39.66\%)         & 0.0685 (17.66\%)        & 0.0447 (12.08\%)        & 0.1057 (26.30\%)        & 0.0648 (30.86\%)        & -                        & -                         \\
\textbf{LKGR}       & 0.0883 (29.11\%)         & 0.0675 (68.00\%)         & 0.0679 (18.70\%)        & 0.0438 (14.38\%)        & 0.1033 (29.24\%)        & 0.0612 (38.56\%)        & -                        & -                         \\ 
\textbf{KGNN-LS}    & 0.0880 (29.55\%)         & 0.0642 (76.64\%)         & 0.0671 (20.12\%)        & 0.0442 (13.35\%)        & 0.1039 (28.49\%)        & 0.0557 (52.24\%)        & 0.1362 (28.63\%)        & 0.056 (70.89\%)          \\
\textbf{KGAT}       & 0.0873 (30.58\%)         & 0.0744 (52.42\%)         & 0.0705 (14.33\%)        & 0.0463 (8.21\%)         & 0.1030 (29.61\%)        & 0.0627 (35.25\%)        & 0.1487 (17.82\%)        & 0.0799 (19.77\%)         \\
\textbf{CKAN}       & 0.0812 (40.39\%)         & 0.0660 (71.82\%)         & 0.0646 (24.77\%)        & 0.0441 (13.61\%)        & 0.0970 (37.64\%)        & 0.0509 (66.60\%)        & 0.1442 (21.50\%)        & 0.0698 (37.11\%)         \\
\textbf{KGIN}       & 0.0978 (16.56\%)         & 0.0848 (33.73\%)         & 0.0698 (15.47\%)        & 0.0451 (11.09\%)        & 0.1147 (16.39\%)        & 0.0716 (18.44\%)        & \uline{0.1687 (3.85\%)} & \uline{0.0915 (4.59\%)}  \\
\textbf{HAKG}       & \uline{0.1008 (13.10\%)} & \uline{0.0931 (21.80\%)} & \uline{0.0778 (3.60\%)} & \uline{0.0501 (0.00\%)} & \uline{0.1319 (1.21\%)} & \uline{0.0848 (0.00\%)} & 0.1642 (6.70\%)          &  0.0907 (5.51\%)            \\ 
\hline
\textbf{{\model}}       & \textbf{0.1140}           & \textbf{0.1134}           & \textbf{0.0806}          & \textbf{0.0501}          & \textbf{0.1335}          & \textbf{0.0848}          & \textbf{0.1752}          & \textbf{0.0957}           \\
\bottomrule
\end{tabular}}
\end{table*}

\subsubsection{Baselines.}
To demonstrate the effectiveness of our proposed model, we compare the overall performance of {\model} with the state-of-the-art methods which include the KG-free method (MF), embedding-based methods (CKE, UGRec, and Hyper-Know), and propagation-based methods (LKGR, KGNN-LS, KGAT, CKAN, KGIN, and HAKG):
\begin{itemize}
\item \textbf{MF}~\cite{BPR} (Matrix factorization) is a benchmark factorization model, which only considers user-item interaction information without using KG. 

\item \textbf{CKE}~\cite{CKE} adopts TransR~\cite{transR} to encode the items’ semantic information and further incorporate it into the MF framework as item representation with the knowledge graph. KG relations are only used as the constraints in TransR training process.

\item \textbf{UGRec}~\cite{UGRec} is an embedding method that integrates the directed relations in KG and the undirected item-item co-occurrence relations simultaneously. While such undirected relations are unavailable for the datasets, the connectivities between items that are co-interacted by the same user are added as the co-occurrence relation.

\item \textbf{Hyper-Know}~\cite{Hyper-Know} is an embedding-based method that embeds the knowledge graph in a hyperbolic space.

\item \textbf{LKGR}~\cite{LKGR} is a hyperbolic GNN-based method with Lorentz model. It uses different information propagation strategies to encode heterogeneous information.

\item \textbf{KGNN-LS}~\cite{KGNN-LS} is a GNN-based model,  which transforms KG into user-specific graphs and enriches item embeddings with GNN and label smoothness regularization.

\item \textbf{KGAT}~\cite{KGAT} is a GNN-based model, which iteratively applies an attentive message-passing scheme over the user-item-entity graph to generate user and item representations.

\item \textbf{CKAN}~\cite{CKAN} is based on KGNN-LS utilizing different neighborhood aggregation schemes on the user-item graph and KG respectively, to integrate the collaborative filtering representation space with the knowledge graph embedding.

\item \textbf{KGIN}~\cite{KGIN} is a GNN-based method to identify the latent intention of users with the interaction history, and performs GNN on the proposed user-intent-item-entity graph.

\item \textbf{HAKG}~\cite{HAKG} is a recently proposed GNN model that captures the underlying hierarchical structure of data in hyperbolic space.
\end{itemize}

\subsubsection{Parameter Settings.}
We implement the proposed {\model} in Pytorch and train it using an RTX 3090 GPU. Our model is optimized with the Adam optimizer, with a fixed batch size of 256 for all three training steps and a negative sample size of 256 as well. The learning rate is initially set to $10^{-4}$, transitioning to $2*10^{-5}$ and $4*10^{-6}$ when the training steps reach 50\% and 75\% of the maximum training steps, respectively. The number of training epochs is 100, 100, and 30 for the three training processes, respectively (except for the first training epoch for Amazon-Book, which is set to 8 due to the significantly larger number of triplets compared to other datasets). The embedding dimension is set to 512, and the margin $\gamma$ in Equation~\ref{loss} is set to 12. Furthermore, we employ an early stopping strategy when the recall@20 does not increase for two consecutive epochs. The baseline settings remain consistent with those in KGIN~\cite{KGIN} and HAKG~\cite{HAKG}.

\subsection{RQ1: Overall Performance}
We present the overall performance on all datasets in Table~\ref{table:results}, including recall@20 and ndcg@20, with the strongest baseline marked by an underline. In addition to the results, we also indicate the relative improvement of our model over each baseline. The baseline results of Last-FM, Yelp2018, and Alibaba-iFashion are taken from HAKG~\cite{HAKG}, while the results of Amazon-Book are from KGIN~\cite{KGIN}, except for HAKG's results on Amazon-Book, which are obtained using HAKG's open-source code. Our findings are as follows:

Compared to all the baselines, our model demonstrates superior performance across the four datasets, outperforming the strongest baseline with respect to recall@20 by 13.10\%, 3.6\%, 1.21\%, and 3.85\% in Last-FM, Yelp2018, Alibaba-iFashion, and Amazon-Book, respectively. This highlights the effectiveness of {\model}. We attribute the improvements to (1) The box and point embeddings representing tags and items in the KG being more suitable for characterizing their connectivity in recommendation scenarios, resulting in a more reasonable distribution in vector space. In contrast, baselines that only use point embedding to model items and tags cannot clearly reflect this subordination. From another perspective, the relative positions of boxes and points also indicate the hierarchy of the entities in KG. In fact, HAKG, which considers the hierarchical relation of the KG, is generally closest to our method, yet still weaker, possibly because: (2) Concept combination more explicitly captures the complexity and diversity of user interests, making our model more expressive in capturing user interests than models that do not consider the connectivity of KG information and user interest in an explicit way.

Upon thoroughly analyzing the performance of {\model} across four distinct datasets, we observe that the improvement on Last-FM is more pronounced than on the other three datasets. Referring to the triplet type ratio in Table~\ref{table:datasets}, we believe this is due to the proportion of \textit{IRT}-type triplets in Last-FM being the highest compared to other datasets. Besides utilizing \textit{TRT} and \textit{IRI} triplets in the first step, the model only employs \textit{IRT} triplets in the subsequent two training processes. Simultaneously, we consider this an advantage of our model, as {\model} has a lower demand for the completeness of the knowledge graph. \textit{IRT} triplets are typically more crucial to train our model, so we do not require as much information about \textit{IRI} and \textit{TRT} triplets as other models, while GNN-based models depend on the quality of the entire KG information. A more in-depth analysis of the influence of triplet type will be discussed in RQ2.

In {\model}, when obtaining the intersection box of an item, it can be considered as utilizing the one-hop neighbor information of the item. Compared to GNN-based models (i.e., KGAT and CKAN), although they claim to aggregate multi-hop neighbor information, their performance remains inferior to ours. Even when compared with other embedding-based baselines, such as CKE (focusing on one-hop neighbor information), the results are still not superior across all datasets. This may suggest that the importance of one-hop neighbors in KG plays a more crucial role than multi-hop nodes in the recommendation scenario, and devoting more attention to one-hop neighbors could benefit the outcomes.

\begin{table*}[t]
\setlength\tabcolsep{1.75pt}
\centering
\caption{Impact of each training step. For the basic pretraining step, we consider removing the whole step and some types of triplets, termed w/o B and only IRT; For the box intersection step, we consider removing the whole step and using a different intersection strategy, termed w/o I and M-M I; For the recommendation step, we consider removing the previous two steps, and the impact of user-bias intersection, termed w/o B\&I, w/o userI, and only userI.  The best results are highlighted in bold. The numbers in brackets indicate the relative improvements of our proposed method over the validation results.}
\label{table:analysis}
\resizebox{\textwidth}{!}{ 
\begin{tabular}{ccccccccc} 
\toprule
                & \multicolumn{2}{c}{\textbf{Last-FM}} & \multicolumn{2}{c}{\textbf{Yelp2018}} & \multicolumn{2}{c}{\textbf{Alibaba-iFashion}} & \multicolumn{2}{c}{\textbf{Amazon-book}}  \\ 
\hline
                & \textbf{recall}  & \textbf{ndcg}     & \textbf{recall} & \textbf{ndcg}       & \textbf{recall} & \textbf{ndcg}               & \textbf{recall} & \textbf{ndcg}           \\ 
\midrule
\textbf{w/o B}  & 0.1092(4.40\%)   & 0.1090(4.04\%)    & 0.0796(1.26\%)  & 0.0500(0.20\%)      & 0.1276(4.62\%)  & 0.0809(4.82\%)              & 0.1733(1.10\%)  & 0.0946(1.16\%)          \\
\textbf{only IRT}   & 0.1084(5.17\%)   & 0.1077(5.29\%)    & 0.0803(0.37\%)  & 0.0501(0.00\%)      & 0.1278(4.46\%)  & 0.0809(4.82\%)              & 0.1725(1.57\%)  & 0.0943(1.48\%)          \\ 
\hline
\textbf{w/o I}  & 0.1069(6.64\%)   & 0.1063(6.68\%)    & 0.0779(3.47\%)  & 0.0488(2.66\%)      & 0.1274(4.79\%)  & 0.0805(5.34\%)              & 0.1665(5.23\%)  & 0.0907(5.51\%)          \\
\textbf{M-M I}  & 0.1079(5.65\%)   & 0.1072(5.78\%)    & 0.0799(0.88\%)  & 0.0500(0.20\%)      & 0.1277(4.54\%)  & 0.0809(4.82\%)              & 0.1722(1.74\%)  & 0.0943(1.48\%)          \\ 
\hline
\textbf{w/o B\&I} & 0.0363(214.05\%) & 0.0370(206.49\%)  & 0.0602(33.89\%) & 0.0384(30.47\%)     & 0.0684(95.18\%) & 0.0397(113.6\%)             & 0.1059(65.44\%) & 0.0543(76.24\%)         \\
\textbf{w/o userI}   & 0.1114(2.33\%)   & 0.1104(2.72\%)    & 0.0795(1.38\%)  & 0.0494(1.42\%)      & 0.1280(4.30\%)  & 0.0810(4.69\%)              & 0.1737(0.86\%)  & 0.0954(0.31\%)          \\
\textbf{only userI}   & 0.0621(83.57\%)  & 0.0620(82.90\%)   & 0.0738(9.21\%)  & 0.0466(7.51\%)      & 0.0920(45.11\%) & 0.0557(52.24\%)             & 0.1479(18.46\%) & 0.0880(8.75\%)          \\ 
\hline
\textbf{Base}       & \textbf{0.1140}  & \textbf{0.1134}   & \textbf{0.0806} & \textbf{0.0501}     & \textbf{0.1335} & \textbf{0.0848}             & \textbf{0.1752} & \textbf{0.0957}         \\
\bottomrule
\end{tabular}}
\end{table*}

\subsection{RQ2: Model Analysis}
In this section, we conduct experiments to analyze the effectiveness of each training step and various components of the proposed model. The experimental results are summarized in Table~\ref{table:analysis}, categorized by their corresponding training stages. For each experiment, we also show the relative performance improvement compared to the base model configuration to demonstrate the contribution of different components. Specifically, we incrementally validate the effects of the Data, different model functions, and training methods. Through these controlled experiments, we aim to provide an in-depth analysis of our multi-stage training methodology and reveal insights into which components are most critical for achieving strong performance. The analysis would guide us toward the most efficient model design by balancing model accuracy and training costs.

\subsubsection{Analysis to the Basic Pretraining Step.}
We initially analyze the role the basic pretraining step plays in the entire training process, which encompasses two perspectives: (1) Is the entire training step crucial? (2) The primary results indicate that the \textit{IRT} triplet may be more important; then, what will the results be if the other types of triplets are removed from the training stage? The first experiment eliminates this entire training step, i.e., training the model starting from the box intersection step, referred to as \textbf{w/o B} (without the Basic pretraining step). The second experiment excludes the \textit{TRT} and \textit{IRI} triplets during the first training step, referred to as \textbf{only IRT}.
\begin{itemize}

\item The outcome achieved by removing the first step is marginally lower than the \textit{base} results. We believe this is because the objectives of the first and second steps are similar. Although the learning targets differ, both steps attempt to obtain suitable point and box representations. Without the basic pretraining step, the box intersection step could still acquire appropriate embeddings.

\item In the first step, removing \textit{TRT} and \textit{IRI} triplets has minimal impact on the results. Also, the subsequent two steps do not utilize \textit{TRT} and \textit{IRI} triplets. This illustrates that the \textit{IRT} triplets in the knowledge graph may play a more direct and crucial role in the recommendation task than the other two types of triplets, and also offers a novel perspective on how to utilize KG rather than incorporating all multi-hop information. Moreover, it also enables our model to have lower requirements for KG triplets, since we only need \textit{ITR} triplets and can still achieve robust performance.

\end{itemize}

\subsubsection{Analysis to the Box Intersection.}
For the second step of training, i.e., box intersection, we are also concerned about two issues: (1) What effect does removing the second step have on the results, termed \textbf{w/o I} (without box Intersection step)? (2) In the \textit{base} experiment, we employ the attention network intersection method. What if we utilize the max-min intersection, termed \textbf{M-M I}.
\begin{itemize}

\item The \textit{w/o I} result is also close to the \textit{base} result, but compared with \textit{w/o B}, it is slightly worse on all datasets. We attribute this difference to the fact that although the first and second steps can both obtain appropriate representations for entities and relations, the first step does not explicitly constrain that the item should be located in the intersection box, while the user interest box is based on the item intersection box. Positioning the item in each individual box, akin to the objective in the first step, may have similar requirements, but it is not direct enough. This also validates the rationality of concept combination and box intersection.

\item Performing the intersection in a purely mathematical manner yields results close to the \textit{base} experiment, which confirms the theoretical rationality of the box intersection operation. The neural network's improved adaptability to data noise and deviation may contribute to the superiority of \textit{base} results over \textit{M-M I}.

\end{itemize}

\subsubsection{Analysis to the Third step.}
In the analysis of the third step, the recommendation step, we conduct three experiments: (1) Remove the first and second steps, and only train the model with the third step, termed \textbf{w/o B\&I} (without Basic pretraining and Intersection); (2) When obtaining the user interest box, only use the intersection box acquired in the second step rather than adding the user-bias intersection box, termed \textbf{w/o userI}; (3) Only use user-bias intersection box in this step, termed \textbf{only userI}.
\begin{itemize}

\item Unlike the results of \textit{w/o I} and \textit{w/o B}, when the first and second steps are removed simultaneously, the performance significantly declines, demonstrating that the model does learn the positional connectivity of tags and items in the previous two steps. The results of \textit{w/o B}, \textit{w/o I}, and \textit{w/o B\&I} indicate that the first or second steps are not required simultaneously, but at least one of them is necessary.

\item In the third step, using only the intersection box embedding obtained in the second step results in a slight performance decrease, suggesting that learning in the second step is effective. However, employing solely the user-bias intersection box leads to a substantial reduction in performance, as the network for user-bias intersection has not been trained in previous steps. Although items and tags have been adequately represented following prior steps, the absence of a training objective makes the intersection operation challenging to train at this stage. Moreover, the superior performance of \textit{only userI} compared to \textit{w/o B\&I} demonstrates that the pre-trained representation benefits the results in the recommendation step.

\end{itemize}

\subsection{R3: Distribution Visualization}
\begin{figure}[h]
\centering
  \includegraphics[scale=1.05]{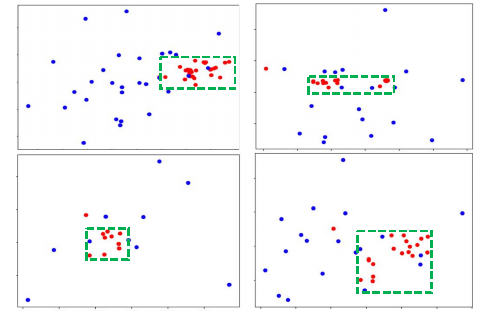}
  \caption{Four cases from \textit{Last-FM}. The red points are the items connected to the relation-tag pair, and the blue points are randomly sampled items.}
  \label{fig:case}
\end{figure}

In this experiment, we investigate whether the final representations of items learned by our model align with their inherent conceptual characteristics defined in the knowledge graph (KG). Specifically, we randomly sample four relation-tag pairs, representing four concepts, and retrieve all items associated with these concept pairs from the Last-FM dataset, along with an equal number of randomly selected items that are not linked to the concepts. To enable visualization, we reduce the high-dimensional item embeddings into two-dimensional points using principal component analysis (PCA). The distribution visualization is shown in Figure~\ref{fig:case}, with each subplot representing an example set of item points corresponding to one concept. We expect items naturally belonging to the same concept to form clusters after projection to 2D space. On the contrary, unrelated negative items should be distributed randomly across the space without clear clustering patterns. By qualitative examination of the visualization, we verify whether our model successfully encodes the semantic correlations indicated in the KG into the item representations. The result demonstrates the effectiveness of our approach in aligning the learned representations to conceptual knowledge.

As depicted in Figure~\ref{fig:case}, items related to the concept (shown in red color) are clustered together, demonstrating that they share similar latent characteristics as defined by the relation-tag pair in the knowledge graph. On the other hand, the randomly selected unrelated items (in blue color) are dispersed arbitrarily across the 2D projected space without a clear pattern. This key observation indicates that the representations learned by our proposed model can effectively adapt to and encode the rich structure and attribute information from the knowledge graph into the embeddings for recommendation. By explicitly modeling the relationships between items and concepts conveyed in the graph, our approach is capable of capturing the underlying semantic correlations and similarity between items possessing certain attributes. As a result, items that fall into the same conceptual categories or share common descriptive traits will have similar vector representations. While unrelated items would be distributed randomly regardless of their conceptual relations. The clear clustering effect aligns well with the inherent item-concept associations, which reveals that our knowledge-enhanced recommendation model can better uncover meaningful item relationships. By harnessing extra knowledge, our approach improves representation learning and leads to more accurate and interpretable recommendations.

\section{RELATED WORK}
Existing KG-enhanced recommendation approaches can be broadly categorized into several main types of methods, each aiming to address key challenges in understanding user preferences and generating personalized recommendations. Some major categories include incorporating knowledge graph embeddings into recommender systems, propagating user preferences along knowledge graph relations, injecting knowledge graph information directly into model architectures and loss functions, generating explanations for recommendations using knowledge graphs, and building conversational recommender systems that interactively construct user profiles. Overall, knowledge graphs have emerged as a powerful technique to enhance many different types of recommender systems by overcoming limitations like cold-start and improving explainability.

Path-based methods~\cite{PLR, NIM, RippleNet, KGCN, PER, KHGT, Meta-path} are one such category that focuses on extracting paths linking target user and item nodes through KG entities. By employing these paths, these methods attempt to anticipate user interests using various strategies, such as recurrent neural networks~\cite{KPRN} or attention mechanisms~\cite{Meta-path}.
PER~\cite{PER}, for example, uncovers meta-path-based latent features to effectively portray the connectivity between users and items. In contrast, KPRN~\cite{KPRN} formulates path representations by integrating the semantics of both entities and relations, thereby inferring the underlying high-order relation of a user-item interaction. To address the considerable number of paths between two nodes, KPRN establishes meta-path patterns as a means to constrain the paths.
RippleNet~\cite{RippleNet}, another prominent method, automatically and iteratively expands a user's potential interests along links in KG. Resorting to a brute-force search method can readily result in labor-intensive and time-consuming feature engineering. Relying on meta-path patterns to filter path instances depends on domain-specific knowledge and human input, which hinders generalization to other contexts and application domains.
In essence, path-based methods focus on utilizing paths connecting user and item nodes in KGs to forecast user interests effectively. Although these methods can be effective in certain scenarios, they frequently necessitate substantial feature engineering, domain-specific knowledge, or human involvement. As a result, these limitations may potentially constrain their scalability and applicability across diverse and evolving scenarios.

Graph neural network (GNN)-based approaches~\cite{LKGR, KGNN-LS, KGAT, CKAN, JKRGCN, KGIN, HAKG} incorporate information from neighboring nodes to refine the representations of central nodes. By executing such propagation recursively, information from multi-hop nodes is ultimately encoded within the final representation, providing a comprehensive understanding of the underlying structure.
For instance, KGAT~\cite{KGAT} iteratively propagates information throughout the KG, fusing graph convolution with an attention mechanism to generate high-quality node representations that effectively capture the complex interactions within the graph.
KGIN~\cite{KGIN}, on the other hand, discerns user intent through KG relations and separates user preferences from intents, thereby enhancing the interpretability of the model's recommendations.
HAKG~\cite{HAKG} represents users, items, entities, and relations within hyperbolic space, employing a hyperbolic aggregation scheme to accumulate relational context across the KG, which allows for better representation of hierarchical structures.
While GNN-based models have demonstrated effectiveness in various contexts, their performance is often contingent upon the quality of the underlying knowledge graph. In real-world scenarios, KGs are frequently sparse and noisy, which imposes limitations on their performance. Consequently, these models may struggle to maintain their efficacy when faced with imperfect or incomplete information, highlighting the need for robust methods that can adapt to such challenges.

Embedding-based approaches~\cite{KTUP, HKB, CKE, DKN, RCF, Hyper-Know, UGRec} typically incorporate KG embedding techniques (e.g., TransE~\cite{transE} and TransR~\cite{transR}) to capture the underlying structure of KGs, while employing additional KG loss to guide the learning process of recommender models. CKE~\cite{CKE} leverages TransR to simultaneously learn latent representations in conjunction with items' semantic representations derived from the knowledge graph. KTUP~\cite{KTUP} utilizes TransH~\cite{transH} on the intersection of user-item and knowledge graph, allowing for a more comprehensive understanding of user preferences.
Hyper-Know~\cite{Hyper-Know} embeds the knowledge graph within hyperbolic space, thereby facilitating the learning of the hierarchical structure inherent in KGs, while designing an adaptive regularization mechanism to modulate item representations. This approach enables better representation of complex and hierarchical relationships, improving the overall effectiveness of the recommendation system.
By integrating KG embeddings and additional loss functions, embedding-based approaches can effectively capture the complex relationships within KGs and provide more accurate and meaningful recommendations. However, these methods still rely on the quality and completeness of the underlying knowledge graph, which can limit their performance in real-world scenarios where KGs may be sparse or noisy.

We concentrate on representing user interests more aptly rather than as a mere point. Our proposed model refer to the concept of previous embedding-based approaches by learning box embeddings~\cite{word2box, BoxE, Q2B} for user interests. Box embeddings are a technique to represent a conjunction of concepts as the intersection of their embedding boxes. For example, the concept "cat pet" can be represented as the box formed by intersecting the embedding boxes for "cat" and "pet". Grounded in the notion that the conjunction of distinct concepts (e.g., relation-tag pairs in the KG) can convey user interests, we utilize the intersection of tag box embeddings to express user preferences. This allows us to represent user interests not as a single point, but as a rich box region covering the latent dependencies among user-tag-item relationships. Consequently, user preferences are explicitly embedded as a box that encapsulates the semantic connections between the tags the user is interested in and the items they interact with.

\section{CONCLUSION}
In this paper, we have presented a novel knowledge-aware recommendation model {\model}, which introduces box representations to effectively encode the information from knowledge graphs. By modeling tags and relations as boxes geometrically, our approach enables more intuitive modeling of the connectivity and complex associations between item sets sharing common attributes. Moreover, we propose to represent each user’s diverse interests as combinations of varying fundamental concept boxes via an intersection operation.
Extensive experiments on four real-world datasets have demonstrated the effectiveness of {\model}, significantly outperforming state-of-the-art baselines in various settings. The in-depth quantitative analysis provides insights into the utility of different types of semantic knowledge relations, shedding light on what kind of information is most valuable for enhancing recommendations. The visualization also shows our model can produce more interpretable recommendations by uncovering the alignments between user preference and item attributes.
In summary, this paper makes important progress in bridging knowledge graphs with advanced recommendation models. The box representations offer a flexible geometric approach to encode rich structured knowledge into embeddings. By explicitly modeling semantic connections and leveraging contextual concept information, knowledge graph-enhanced recommendations can better capture user interests and item characteristics to provide more accurate, diverse, and interpretable recommendation results.

\begin{acks}
This work is founded by National Natural Science Foundation of China (NSFC62306276/NSFCU23B2055/NSFCU19B2027/NSFC91846204), Zhejiang Provincial Natural Science Foundation of China (No. LQ23F020017), Ningbo Natural Science Foundation (2023J291), Yongjiang Talent Introduction Programme (2022A-238-G),  Fundamental Research Funds for the Central Universities (226-2023-00138).
\end{acks}


\bibliographystyle{ACM-Reference-Format}
\bibliography{sample}


\begin{thebibliography}{35}


\ifx \showCODEN    \undefined \def \showCODEN     #1{\unskip}     \fi
\ifx \showDOI      \undefined \def \showDOI       #1{#1}\fi
\ifx \showISBNx    \undefined \def \showISBNx     #1{\unskip}     \fi
\ifx \showISBNxiii \undefined \def \showISBNxiii  #1{\unskip}     \fi
\ifx \showISSN     \undefined \def \showISSN      #1{\unskip}     \fi
\ifx \showLCCN     \undefined \def \showLCCN      #1{\unskip}     \fi
\ifx \shownote     \undefined \def \shownote      #1{#1}          \fi
\ifx \showarticletitle \undefined \def \showarticletitle #1{#1}   \fi
\ifx \showURL      \undefined \def \showURL       {\relax}        \fi
\providecommand\bibfield[2]{#2}
\providecommand\bibinfo[2]{#2}
\providecommand\natexlab[1]{#1}
\providecommand\showeprint[2][]{arXiv:#2}

\bibitem[\protect\citeauthoryear{Abboud, Ceylan, Lukasiewicz, and Salvatori}{Abboud et~al\mbox{.}}{2020}]%
        {BoxE}
\bibfield{author}{\bibinfo{person}{Ralph Abboud}, \bibinfo{person}{{\.I}smail~{\.I}lkan Ceylan}, \bibinfo{person}{Thomas Lukasiewicz}, {and} \bibinfo{person}{Tommaso Salvatori}.} \bibinfo{year}{2020}\natexlab{}.
\newblock \showarticletitle{BoxE: {A} Box Embedding Model for Knowledge Base Completion}. In \bibinfo{booktitle}{\emph{Advances in Neural Information Processing Systems 33: Annual Conference on Neural Information Processing Systems 2020, NeurIPS 2020, December 6-12, 2020, virtual}}, \bibfield{editor}{\bibinfo{person}{Hugo Larochelle}, \bibinfo{person}{Marc'Aurelio Ranzato}, \bibinfo{person}{Raia Hadsell}, \bibinfo{person}{Maria{-}Florina Balcan}, {and} \bibinfo{person}{Hsuan{-}Tien Lin}} (Eds.).
\newblock


\bibitem[\protect\citeauthoryear{Ai, Azizi, Chen, and Zhang}{Ai et~al\mbox{.}}{2018}]%
        {HKB}
\bibfield{author}{\bibinfo{person}{Qingyao Ai}, \bibinfo{person}{Vahid Azizi}, \bibinfo{person}{Xu Chen}, {and} \bibinfo{person}{Yongfeng Zhang}.} \bibinfo{year}{2018}\natexlab{}.
\newblock \showarticletitle{Learning Heterogeneous Knowledge Base Embeddings for Explainable Recommendation}.
\newblock \bibinfo{journal}{\emph{Algorithms}} \bibinfo{volume}{11}, \bibinfo{number}{9} (\bibinfo{year}{2018}), \bibinfo{pages}{137}.
\newblock
\urldef\tempurl%
\url{https://doi.org/10.3390/a11090137}
\showDOI{\tempurl}


\bibitem[\protect\citeauthoryear{Bollacker, Evans, Paritosh, Sturge, and Taylor}{Bollacker et~al\mbox{.}}{2008}]%
        {Freebase}
\bibfield{author}{\bibinfo{person}{Kurt~D. Bollacker}, \bibinfo{person}{Colin Evans}, \bibinfo{person}{Praveen~K. Paritosh}, \bibinfo{person}{Tim Sturge}, {and} \bibinfo{person}{Jamie Taylor}.} \bibinfo{year}{2008}\natexlab{}.
\newblock \showarticletitle{Freebase: a collaboratively created graph database for structuring human knowledge}. In \bibinfo{booktitle}{\emph{Proceedings of the {ACM} {SIGMOD} International Conference on Management of Data, {SIGMOD} 2008, Vancouver, BC, Canada, June 10-12, 2008}}, \bibfield{editor}{\bibinfo{person}{Jason~Tsong{-}Li Wang}} (Ed.). \bibinfo{publisher}{{ACM}}, \bibinfo{pages}{1247--1250}.
\newblock
\urldef\tempurl%
\url{https://doi.org/10.1145/1376616.1376746}
\showDOI{\tempurl}


\bibitem[\protect\citeauthoryear{Bordes, Usunier, Garc{\'{\i}}a{-}Dur{\'{a}}n, Weston, and Yakhnenko}{Bordes et~al\mbox{.}}{2013}]%
        {transE}
\bibfield{author}{\bibinfo{person}{Antoine Bordes}, \bibinfo{person}{Nicolas Usunier}, \bibinfo{person}{Alberto Garc{\'{\i}}a{-}Dur{\'{a}}n}, \bibinfo{person}{Jason Weston}, {and} \bibinfo{person}{Oksana Yakhnenko}.} \bibinfo{year}{2013}\natexlab{}.
\newblock \showarticletitle{Translating Embeddings for Modeling Multi-relational Data}. In \bibinfo{booktitle}{\emph{Advances in Neural Information Processing Systems 26: 27th Annual Conference on Neural Information Processing Systems 2013. Proceedings of a meeting held December 5-8, 2013, Lake Tahoe, Nevada, United States}}, \bibfield{editor}{\bibinfo{person}{Christopher J.~C. Burges}, \bibinfo{person}{L{\'{e}}on Bottou}, \bibinfo{person}{Zoubin Ghahramani}, {and} \bibinfo{person}{Kilian~Q. Weinberger}} (Eds.). \bibinfo{pages}{2787--2795}.
\newblock


\bibitem[\protect\citeauthoryear{Cao, Wang, He, Hu, and Chua}{Cao et~al\mbox{.}}{2019}]%
        {KTUP}
\bibfield{author}{\bibinfo{person}{Yixin Cao}, \bibinfo{person}{Xiang Wang}, \bibinfo{person}{Xiangnan He}, \bibinfo{person}{Zikun Hu}, {and} \bibinfo{person}{Tat{-}Seng Chua}.} \bibinfo{year}{2019}\natexlab{}.
\newblock \showarticletitle{Unifying Knowledge Graph Learning and Recommendation: Towards a Better Understanding of User Preferences}. In \bibinfo{booktitle}{\emph{The World Wide Web Conference, {WWW} 2019, San Francisco, CA, USA, May 13-17, 2019}}, \bibfield{editor}{\bibinfo{person}{Ling Liu}, \bibinfo{person}{Ryen~W. White}, \bibinfo{person}{Amin Mantrach}, \bibinfo{person}{Fabrizio Silvestri}, \bibinfo{person}{Julian~J. McAuley}, \bibinfo{person}{Ricardo Baeza{-}Yates}, {and} \bibinfo{person}{Leila Zia}} (Eds.). \bibinfo{publisher}{{ACM}}, \bibinfo{pages}{151--161}.
\newblock
\urldef\tempurl%
\url{https://doi.org/10.1145/3308558.3313705}
\showDOI{\tempurl}


\bibitem[\protect\citeauthoryear{Catherine and Cohen}{Catherine and Cohen}{2016}]%
        {PLR}
\bibfield{author}{\bibinfo{person}{Rose Catherine} {and} \bibinfo{person}{William~W. Cohen}.} \bibinfo{year}{2016}\natexlab{}.
\newblock \showarticletitle{Personalized Recommendations using Knowledge Graphs: {A} Probabilistic Logic Programming Approach}. In \bibinfo{booktitle}{\emph{Proceedings of the 10th {ACM} Conference on Recommender Systems, Boston, MA, USA, September 15-19, 2016}}, \bibfield{editor}{\bibinfo{person}{Shilad Sen}, \bibinfo{person}{Werner Geyer}, \bibinfo{person}{Jill Freyne}, {and} \bibinfo{person}{Pablo Castells}} (Eds.). \bibinfo{publisher}{{ACM}}, \bibinfo{pages}{325--332}.
\newblock
\urldef\tempurl%
\url{https://doi.org/10.1145/2959100.2959131}
\showDOI{\tempurl}


\bibitem[\protect\citeauthoryear{Chen, Yang, Zhang, Zhao, Meng, Hao, and King}{Chen et~al\mbox{.}}{2022}]%
        {LKGR}
\bibfield{author}{\bibinfo{person}{Yankai Chen}, \bibinfo{person}{Menglin Yang}, \bibinfo{person}{Yingxue Zhang}, \bibinfo{person}{Mengchen Zhao}, \bibinfo{person}{Ziqiao Meng}, \bibinfo{person}{Jianye Hao}, {and} \bibinfo{person}{Irwin King}.} \bibinfo{year}{2022}\natexlab{}.
\newblock \showarticletitle{Modeling Scale-free Graphs with Hyperbolic Geometry for Knowledge-aware Recommendation}. In \bibinfo{booktitle}{\emph{{WSDM} '22: The Fifteenth {ACM} International Conference on Web Search and Data Mining, Virtual Event / Tempe, AZ, USA, February 21 - 25, 2022}}, \bibfield{editor}{\bibinfo{person}{K.~Selcuk Candan}, \bibinfo{person}{Huan Liu}, \bibinfo{person}{Leman Akoglu}, \bibinfo{person}{Xin~Luna Dong}, {and} \bibinfo{person}{Jiliang Tang}} (Eds.). \bibinfo{publisher}{{ACM}}, \bibinfo{pages}{94--102}.
\newblock
\urldef\tempurl%
\url{https://doi.org/10.1145/3488560.3498419}
\showDOI{\tempurl}


\bibitem[\protect\citeauthoryear{Dasgupta, Boratko, Mishra, Atmakuri, Patel, Li, and McCallum}{Dasgupta et~al\mbox{.}}{2022}]%
        {word2box}
\bibfield{author}{\bibinfo{person}{Shib Dasgupta}, \bibinfo{person}{Michael Boratko}, \bibinfo{person}{Siddhartha Mishra}, \bibinfo{person}{Shriya Atmakuri}, \bibinfo{person}{Dhruvesh Patel}, \bibinfo{person}{Xiang Li}, {and} \bibinfo{person}{Andrew McCallum}.} \bibinfo{year}{2022}\natexlab{}.
\newblock \showarticletitle{{W}ord2{B}ox: Capturing Set-Theoretic Semantics of Words using Box Embeddings}. In \bibinfo{booktitle}{\emph{Proceedings of the 60th Annual Meeting of the Association for Computational Linguistics (Volume 1: Long Papers)}}. \bibinfo{publisher}{Association for Computational Linguistics}, \bibinfo{address}{Dublin, Ireland}, \bibinfo{pages}{2263--2276}.
\newblock
\urldef\tempurl%
\url{https://doi.org/10.18653/v1/2022.acl-long.161}
\showDOI{\tempurl}


\bibitem[\protect\citeauthoryear{Du, Zhu, Chen, Zheng, and Gao}{Du et~al\mbox{.}}{2022}]%
        {HAKG}
\bibfield{author}{\bibinfo{person}{Yuntao Du}, \bibinfo{person}{Xinjun Zhu}, \bibinfo{person}{Lu Chen}, \bibinfo{person}{Baihua Zheng}, {and} \bibinfo{person}{Yunjun Gao}.} \bibinfo{year}{2022}\natexlab{}.
\newblock \showarticletitle{{HAKG:} Hierarchy-Aware Knowledge Gated Network for Recommendation}. In \bibinfo{booktitle}{\emph{{SIGIR} '22: The 45th International {ACM} {SIGIR} Conference on Research and Development in Information Retrieval, Madrid, Spain, July 11 - 15, 2022}}, \bibfield{editor}{\bibinfo{person}{Enrique Amig{\'{o}}}, \bibinfo{person}{Pablo Castells}, \bibinfo{person}{Julio Gonzalo}, \bibinfo{person}{Ben Carterette}, \bibinfo{person}{J.~Shane Culpepper}, {and} \bibinfo{person}{Gabriella Kazai}} (Eds.). \bibinfo{publisher}{{ACM}}, \bibinfo{pages}{1390--1400}.
\newblock
\urldef\tempurl%
\url{https://doi.org/10.1145/3477495.3531987}
\showDOI{\tempurl}


\bibitem[\protect\citeauthoryear{Gharibshah, Zhu, Hainline, and Conway}{Gharibshah et~al\mbox{.}}{2020}]%
        {DLUI}
\bibfield{author}{\bibinfo{person}{Zhabiz Gharibshah}, \bibinfo{person}{Xingquan Zhu}, \bibinfo{person}{Arthur Hainline}, {and} \bibinfo{person}{Michael Conway}.} \bibinfo{year}{2020}\natexlab{}.
\newblock \showarticletitle{Deep Learning for User Interest and Response Prediction in Online Display Advertising}.
\newblock \bibinfo{journal}{\emph{Data Sci. Eng.}} \bibinfo{volume}{5}, \bibinfo{number}{1} (\bibinfo{year}{2020}), \bibinfo{pages}{12--26}.
\newblock
\urldef\tempurl%
\url{https://doi.org/10.1007/s41019-019-00115-y}
\showDOI{\tempurl}


\bibitem[\protect\citeauthoryear{Hu, Shi, Zhao, and Yu}{Hu et~al\mbox{.}}{2018}]%
        {Meta-path}
\bibfield{author}{\bibinfo{person}{Binbin Hu}, \bibinfo{person}{Chuan Shi}, \bibinfo{person}{Wayne~Xin Zhao}, {and} \bibinfo{person}{Philip~S. Yu}.} \bibinfo{year}{2018}\natexlab{}.
\newblock \showarticletitle{Leveraging Meta-path based Context for Top- {N} Recommendation with {A} Neural Co-Attention Model}. In \bibinfo{booktitle}{\emph{Proceedings of the 24th {ACM} {SIGKDD} International Conference on Knowledge Discovery {\&} Data Mining, {KDD} 2018, London, UK, August 19-23, 2018}}, \bibfield{editor}{\bibinfo{person}{Yike Guo} {and} \bibinfo{person}{Faisal Farooq}} (Eds.). \bibinfo{publisher}{{ACM}}, \bibinfo{pages}{1531--1540}.
\newblock
\urldef\tempurl%
\url{https://doi.org/10.1145/3219819.3219965}
\showDOI{\tempurl}


\bibitem[\protect\citeauthoryear{Hu, Koren, and Volinsky}{Hu et~al\mbox{.}}{2008}]%
        {CF}
\bibfield{author}{\bibinfo{person}{Yifan Hu}, \bibinfo{person}{Yehuda Koren}, {and} \bibinfo{person}{Chris Volinsky}.} \bibinfo{year}{2008}\natexlab{}.
\newblock \showarticletitle{Collaborative Filtering for Implicit Feedback Datasets}. In \bibinfo{booktitle}{\emph{Proceedings of the 8th {IEEE} International Conference on Data Mining {(ICDM} 2008), December 15-19, 2008, Pisa, Italy}}. \bibinfo{publisher}{{IEEE} Computer Society}, \bibinfo{pages}{263--272}.
\newblock
\urldef\tempurl%
\url{https://doi.org/10.1109/ICDM.2008.22}
\showDOI{\tempurl}


\bibitem[\protect\citeauthoryear{Jin, Qin, Fang, Du, Zhang, Yu, Zhang, and Smola}{Jin et~al\mbox{.}}{2020}]%
        {NIM}
\bibfield{author}{\bibinfo{person}{Jiarui Jin}, \bibinfo{person}{Jiarui Qin}, \bibinfo{person}{Yuchen Fang}, \bibinfo{person}{Kounianhua Du}, \bibinfo{person}{Weinan Zhang}, \bibinfo{person}{Yong Yu}, \bibinfo{person}{Zheng Zhang}, {and} \bibinfo{person}{Alexander~J. Smola}.} \bibinfo{year}{2020}\natexlab{}.
\newblock \showarticletitle{An Efficient Neighborhood-based Interaction Model for Recommendation on Heterogeneous Graph}. In \bibinfo{booktitle}{\emph{{KDD} '20: The 26th {ACM} {SIGKDD} Conference on Knowledge Discovery and Data Mining, Virtual Event, CA, USA, August 23-27, 2020}}, \bibfield{editor}{\bibinfo{person}{Rajesh Gupta}, \bibinfo{person}{Yan Liu}, \bibinfo{person}{Jiliang Tang}, {and} \bibinfo{person}{B.~Aditya Prakash}} (Eds.). \bibinfo{publisher}{{ACM}}, \bibinfo{pages}{75--84}.
\newblock
\urldef\tempurl%
\url{https://doi.org/10.1145/3394486.3403050}
\showDOI{\tempurl}


\bibitem[\protect\citeauthoryear{Koren}{Koren}{2008}]%
        {FM}
\bibfield{author}{\bibinfo{person}{Yehuda Koren}.} \bibinfo{year}{2008}\natexlab{}.
\newblock \showarticletitle{Factorization meets the neighborhood: a multifaceted collaborative filtering model}. In \bibinfo{booktitle}{\emph{Proceedings of the 14th {ACM} {SIGKDD} International Conference on Knowledge Discovery and Data Mining, Las Vegas, Nevada, USA, August 24-27, 2008}}, \bibfield{editor}{\bibinfo{person}{Ying Li}, \bibinfo{person}{Bing Liu}, {and} \bibinfo{person}{Sunita Sarawagi}} (Eds.). \bibinfo{publisher}{{ACM}}, \bibinfo{pages}{426--434}.
\newblock
\urldef\tempurl%
\url{https://doi.org/10.1145/1401890.1401944}
\showDOI{\tempurl}


\bibitem[\protect\citeauthoryear{Krichene and Rendle}{Krichene and Rendle}{2020}]%
        {SM}
\bibfield{author}{\bibinfo{person}{Walid Krichene} {and} \bibinfo{person}{Steffen Rendle}.} \bibinfo{year}{2020}\natexlab{}.
\newblock \showarticletitle{On Sampled Metrics for Item Recommendation}. In \bibinfo{booktitle}{\emph{{KDD} '20: The 26th {ACM} {SIGKDD} Conference on Knowledge Discovery and Data Mining, Virtual Event, CA, USA, August 23-27, 2020}}, \bibfield{editor}{\bibinfo{person}{Rajesh Gupta}, \bibinfo{person}{Yan Liu}, \bibinfo{person}{Jiliang Tang}, {and} \bibinfo{person}{B.~Aditya Prakash}} (Eds.). \bibinfo{publisher}{{ACM}}, \bibinfo{pages}{1748--1757}.
\newblock
\urldef\tempurl%
\url{https://doi.org/10.1145/3394486.3403226}
\showDOI{\tempurl}


\bibitem[\protect\citeauthoryear{Lin, Liu, Sun, Liu, and Zhu}{Lin et~al\mbox{.}}{2015}]%
        {transR}
\bibfield{author}{\bibinfo{person}{Yankai Lin}, \bibinfo{person}{Zhiyuan Liu}, \bibinfo{person}{Maosong Sun}, \bibinfo{person}{Yang Liu}, {and} \bibinfo{person}{Xuan Zhu}.} \bibinfo{year}{2015}\natexlab{}.
\newblock \showarticletitle{Learning Entity and Relation Embeddings for Knowledge Graph Completion}. In \bibinfo{booktitle}{\emph{Proceedings of the Twenty-Ninth {AAAI} Conference on Artificial Intelligence, January 25-30, 2015, Austin, Texas, {USA}}}, \bibfield{editor}{\bibinfo{person}{Blai Bonet} {and} \bibinfo{person}{Sven Koenig}} (Eds.). \bibinfo{publisher}{{AAAI} Press}, \bibinfo{pages}{2181--2187}.
\newblock


\bibitem[\protect\citeauthoryear{Ma, Ma, Zhang, Wu, Liu, and Coates}{Ma et~al\mbox{.}}{2021}]%
        {Hyper-Know}
\bibfield{author}{\bibinfo{person}{Chen Ma}, \bibinfo{person}{Liheng Ma}, \bibinfo{person}{Yingxue Zhang}, \bibinfo{person}{Haolun Wu}, \bibinfo{person}{Xue Liu}, {and} \bibinfo{person}{Mark Coates}.} \bibinfo{year}{2021}\natexlab{}.
\newblock \showarticletitle{Knowledge-Enhanced Top-K Recommendation in Poincar{\'{e}} Ball}. In \bibinfo{booktitle}{\emph{Thirty-Fifth {AAAI} Conference on Artificial Intelligence, {AAAI} 2021, Thirty-Third Conference on Innovative Applications of Artificial Intelligence, {IAAI} 2021, The Eleventh Symposium on Educational Advances in Artificial Intelligence, {EAAI} 2021, Virtual Event, February 2-9, 2021}}. \bibinfo{publisher}{{AAAI} Press}, \bibinfo{pages}{4285--4293}.
\newblock


\bibitem[\protect\citeauthoryear{Miller}{Miller}{1995}]%
        {Wordnet}
\bibfield{author}{\bibinfo{person}{George~A Miller}.} \bibinfo{year}{1995}\natexlab{}.
\newblock \showarticletitle{WordNet: a lexical database for English}.
\newblock \bibinfo{journal}{\emph{Commun. ACM}} \bibinfo{volume}{38}, \bibinfo{number}{11} (\bibinfo{year}{1995}), \bibinfo{pages}{39--41}.
\newblock


\bibitem[\protect\citeauthoryear{Ren, Hu, and Leskovec}{Ren et~al\mbox{.}}{2020}]%
        {Q2B}
\bibfield{author}{\bibinfo{person}{Hongyu Ren}, \bibinfo{person}{Weihua Hu}, {and} \bibinfo{person}{Jure Leskovec}.} \bibinfo{year}{2020}\natexlab{}.
\newblock \showarticletitle{Query2box: Reasoning over Knowledge Graphs in Vector Space Using Box Embeddings}. In \bibinfo{booktitle}{\emph{8th International Conference on Learning Representations, {ICLR} 2020, Addis Ababa, Ethiopia, April 26-30, 2020}}. \bibinfo{publisher}{OpenReview.net}.
\newblock


\bibitem[\protect\citeauthoryear{Rendle, Freudenthaler, Gantner, and Schmidt{-}Thieme}{Rendle et~al\mbox{.}}{2009}]%
        {BPR}
\bibfield{author}{\bibinfo{person}{Steffen Rendle}, \bibinfo{person}{Christoph Freudenthaler}, \bibinfo{person}{Zeno Gantner}, {and} \bibinfo{person}{Lars Schmidt{-}Thieme}.} \bibinfo{year}{2009}\natexlab{}.
\newblock \showarticletitle{{BPR:} Bayesian Personalized Ranking from Implicit Feedback}. In \bibinfo{booktitle}{\emph{{UAI} 2009, Proceedings of the Twenty-Fifth Conference on Uncertainty in Artificial Intelligence, Montreal, QC, Canada, June 18-21, 2009}}, \bibfield{editor}{\bibinfo{person}{Jeff~A. Bilmes} {and} \bibinfo{person}{Andrew~Y. Ng}} (Eds.). \bibinfo{publisher}{{AUAI} Press}, \bibinfo{pages}{452--461}.
\newblock


\bibitem[\protect\citeauthoryear{Tian, Yang, Ren, Wang, Wu, Wang, and Li}{Tian et~al\mbox{.}}{2021}]%
        {JKRGCN}
\bibfield{author}{\bibinfo{person}{Yu Tian}, \bibinfo{person}{Yuhao Yang}, \bibinfo{person}{Xudong Ren}, \bibinfo{person}{Pengfei Wang}, \bibinfo{person}{Fangzhao Wu}, \bibinfo{person}{Qian Wang}, {and} \bibinfo{person}{Chenliang Li}.} \bibinfo{year}{2021}\natexlab{}.
\newblock \showarticletitle{Joint Knowledge Pruning and Recurrent Graph Convolution for News Recommendation}. In \bibinfo{booktitle}{\emph{{SIGIR} '21: The 44th International {ACM} {SIGIR} Conference on Research and Development in Information Retrieval, Virtual Event, Canada, July 11-15, 2021}}, \bibfield{editor}{\bibinfo{person}{Fernando Diaz}, \bibinfo{person}{Chirag Shah}, \bibinfo{person}{Torsten Suel}, \bibinfo{person}{Pablo Castells}, \bibinfo{person}{Rosie Jones}, {and} \bibinfo{person}{Tetsuya Sakai}} (Eds.). \bibinfo{publisher}{{ACM}}, \bibinfo{pages}{51--60}.
\newblock
\urldef\tempurl%
\url{https://doi.org/10.1145/3404835.3462912}
\showDOI{\tempurl}


\bibitem[\protect\citeauthoryear{Wang, Zhang, Wang, Zhao, Li, Xie, and Guo}{Wang et~al\mbox{.}}{2018a}]%
        {RippleNet}
\bibfield{author}{\bibinfo{person}{Hongwei Wang}, \bibinfo{person}{Fuzheng Zhang}, \bibinfo{person}{Jialin Wang}, \bibinfo{person}{Miao Zhao}, \bibinfo{person}{Wenjie Li}, \bibinfo{person}{Xing Xie}, {and} \bibinfo{person}{Minyi Guo}.} \bibinfo{year}{2018}\natexlab{a}.
\newblock \showarticletitle{RippleNet: Propagating User Preferences on the Knowledge Graph for Recommender Systems}. In \bibinfo{booktitle}{\emph{Proceedings of the 27th {ACM} International Conference on Information and Knowledge Management, {CIKM} 2018, Torino, Italy, October 22-26, 2018}}, \bibfield{editor}{\bibinfo{person}{Alfredo Cuzzocrea}, \bibinfo{person}{James Allan}, \bibinfo{person}{Norman~W. Paton}, \bibinfo{person}{Divesh Srivastava}, \bibinfo{person}{Rakesh Agrawal}, \bibinfo{person}{Andrei~Z. Broder}, \bibinfo{person}{Mohammed~J. Zaki}, \bibinfo{person}{K.~Sel{\c{c}}uk Candan}, \bibinfo{person}{Alexandros Labrinidis}, \bibinfo{person}{Assaf Schuster}, {and} \bibinfo{person}{Haixun Wang}} (Eds.). \bibinfo{publisher}{{ACM}}, \bibinfo{pages}{417--426}.
\newblock
\urldef\tempurl%
\url{https://doi.org/10.1145/3269206.3271739}
\showDOI{\tempurl}


\bibitem[\protect\citeauthoryear{Wang, Zhang, Xie, and Guo}{Wang et~al\mbox{.}}{2018b}]%
        {DKN}
\bibfield{author}{\bibinfo{person}{Hongwei Wang}, \bibinfo{person}{Fuzheng Zhang}, \bibinfo{person}{Xing Xie}, {and} \bibinfo{person}{Minyi Guo}.} \bibinfo{year}{2018}\natexlab{b}.
\newblock \showarticletitle{{DKN:} Deep Knowledge-Aware Network for News Recommendation}. In \bibinfo{booktitle}{\emph{Proceedings of the 2018 World Wide Web Conference on World Wide Web, {WWW} 2018, Lyon, France, April 23-27, 2018}}, \bibfield{editor}{\bibinfo{person}{Pierre{-}Antoine Champin}, \bibinfo{person}{Fabien Gandon}, \bibinfo{person}{Mounia Lalmas}, {and} \bibinfo{person}{Panagiotis~G. Ipeirotis}} (Eds.). \bibinfo{publisher}{{ACM}}, \bibinfo{pages}{1835--1844}.
\newblock
\urldef\tempurl%
\url{https://doi.org/10.1145/3178876.3186175}
\showDOI{\tempurl}


\bibitem[\protect\citeauthoryear{Wang, Zhang, Zhang, Leskovec, Zhao, Li, and Wang}{Wang et~al\mbox{.}}{2019c}]%
        {KGNN-LS}
\bibfield{author}{\bibinfo{person}{Hongwei Wang}, \bibinfo{person}{Fuzheng Zhang}, \bibinfo{person}{Mengdi Zhang}, \bibinfo{person}{Jure Leskovec}, \bibinfo{person}{Miao Zhao}, \bibinfo{person}{Wenjie Li}, {and} \bibinfo{person}{Zhongyuan Wang}.} \bibinfo{year}{2019}\natexlab{c}.
\newblock \showarticletitle{Knowledge-aware Graph Neural Networks with Label Smoothness Regularization for Recommender Systems}. In \bibinfo{booktitle}{\emph{Proceedings of the 25th {ACM} {SIGKDD} International Conference on Knowledge Discovery {\&} Data Mining, {KDD} 2019, Anchorage, AK, USA, August 4-8, 2019}}, \bibfield{editor}{\bibinfo{person}{Ankur Teredesai}, \bibinfo{person}{Vipin Kumar}, \bibinfo{person}{Ying Li}, \bibinfo{person}{R{\'{o}}mer Rosales}, \bibinfo{person}{Evimaria Terzi}, {and} \bibinfo{person}{George Karypis}} (Eds.). \bibinfo{publisher}{{ACM}}, \bibinfo{pages}{968--977}.
\newblock
\urldef\tempurl%
\url{https://doi.org/10.1145/3292500.3330836}
\showDOI{\tempurl}


\bibitem[\protect\citeauthoryear{Wang, Zhao, Xie, Li, and Guo}{Wang et~al\mbox{.}}{2019d}]%
        {KGCN}
\bibfield{author}{\bibinfo{person}{Hongwei Wang}, \bibinfo{person}{Miao Zhao}, \bibinfo{person}{Xing Xie}, \bibinfo{person}{Wenjie Li}, {and} \bibinfo{person}{Minyi Guo}.} \bibinfo{year}{2019}\natexlab{d}.
\newblock \showarticletitle{Knowledge Graph Convolutional Networks for Recommender Systems}. In \bibinfo{booktitle}{\emph{The World Wide Web Conference, {WWW} 2019, San Francisco, CA, USA, May 13-17, 2019}}, \bibfield{editor}{\bibinfo{person}{Ling Liu}, \bibinfo{person}{Ryen~W. White}, \bibinfo{person}{Amin Mantrach}, \bibinfo{person}{Fabrizio Silvestri}, \bibinfo{person}{Julian~J. McAuley}, \bibinfo{person}{Ricardo Baeza{-}Yates}, {and} \bibinfo{person}{Leila Zia}} (Eds.). \bibinfo{publisher}{{ACM}}, \bibinfo{pages}{3307--3313}.
\newblock
\urldef\tempurl%
\url{https://doi.org/10.1145/3308558.3313417}
\showDOI{\tempurl}


\bibitem[\protect\citeauthoryear{Wang, He, Cao, Liu, and Chua}{Wang et~al\mbox{.}}{2019a}]%
        {KGAT}
\bibfield{author}{\bibinfo{person}{Xiang Wang}, \bibinfo{person}{Xiangnan He}, \bibinfo{person}{Yixin Cao}, \bibinfo{person}{Meng Liu}, {and} \bibinfo{person}{Tat{-}Seng Chua}.} \bibinfo{year}{2019}\natexlab{a}.
\newblock \showarticletitle{{KGAT:} Knowledge Graph Attention Network for Recommendation}. In \bibinfo{booktitle}{\emph{Proceedings of the 25th {ACM} {SIGKDD} International Conference on Knowledge Discovery {\&} Data Mining, {KDD} 2019, Anchorage, AK, USA, August 4-8, 2019}}, \bibfield{editor}{\bibinfo{person}{Ankur Teredesai}, \bibinfo{person}{Vipin Kumar}, \bibinfo{person}{Ying Li}, \bibinfo{person}{R{\'{o}}mer Rosales}, \bibinfo{person}{Evimaria Terzi}, {and} \bibinfo{person}{George Karypis}} (Eds.). \bibinfo{publisher}{{ACM}}, \bibinfo{pages}{950--958}.
\newblock
\urldef\tempurl%
\url{https://doi.org/10.1145/3292500.3330989}
\showDOI{\tempurl}


\bibitem[\protect\citeauthoryear{Wang, Huang, Wang, Yuan, Liu, He, and Chua}{Wang et~al\mbox{.}}{2021}]%
        {KGIN}
\bibfield{author}{\bibinfo{person}{Xiang Wang}, \bibinfo{person}{Tinglin Huang}, \bibinfo{person}{Dingxian Wang}, \bibinfo{person}{Yancheng Yuan}, \bibinfo{person}{Zhenguang Liu}, \bibinfo{person}{Xiangnan He}, {and} \bibinfo{person}{Tat{-}Seng Chua}.} \bibinfo{year}{2021}\natexlab{}.
\newblock \showarticletitle{Learning Intents behind Interactions with Knowledge Graph for Recommendation}. In \bibinfo{booktitle}{\emph{{WWW} '21: The Web Conference 2021, Virtual Event / Ljubljana, Slovenia, April 19-23, 2021}}, \bibfield{editor}{\bibinfo{person}{Jure Leskovec}, \bibinfo{person}{Marko Grobelnik}, \bibinfo{person}{Marc Najork}, \bibinfo{person}{Jie Tang}, {and} \bibinfo{person}{Leila Zia}} (Eds.). \bibinfo{publisher}{{ACM} / {IW3C2}}, \bibinfo{pages}{878--887}.
\newblock
\urldef\tempurl%
\url{https://doi.org/10.1145/3442381.3450133}
\showDOI{\tempurl}


\bibitem[\protect\citeauthoryear{Wang, Wang, Xu, He, Cao, and Chua}{Wang et~al\mbox{.}}{2019b}]%
        {KPRN}
\bibfield{author}{\bibinfo{person}{Xiang Wang}, \bibinfo{person}{Dingxian Wang}, \bibinfo{person}{Canran Xu}, \bibinfo{person}{Xiangnan He}, \bibinfo{person}{Yixin Cao}, {and} \bibinfo{person}{Tat{-}Seng Chua}.} \bibinfo{year}{2019}\natexlab{b}.
\newblock \showarticletitle{Explainable Reasoning over Knowledge Graphs for Recommendation}. In \bibinfo{booktitle}{\emph{The Thirty-Third {AAAI} Conference on Artificial Intelligence, {AAAI} 2019, The Thirty-First Innovative Applications of Artificial Intelligence Conference, {IAAI} 2019, The Ninth {AAAI} Symposium on Educational Advances in Artificial Intelligence, {EAAI} 2019, Honolulu, Hawaii, USA, January 27 - February 1, 2019}}. \bibinfo{publisher}{{AAAI} Press}, \bibinfo{pages}{5329--5336}.
\newblock
\urldef\tempurl%
\url{https://doi.org/10.1609/aaai.v33i01.33015329}
\showDOI{\tempurl}


\bibitem[\protect\citeauthoryear{Wang, Lin, Tan, Chen, and Liu}{Wang et~al\mbox{.}}{2020}]%
        {CKAN}
\bibfield{author}{\bibinfo{person}{Ze Wang}, \bibinfo{person}{Guangyan Lin}, \bibinfo{person}{Huobin Tan}, \bibinfo{person}{Qinghong Chen}, {and} \bibinfo{person}{Xiyang Liu}.} \bibinfo{year}{2020}\natexlab{}.
\newblock \showarticletitle{{CKAN:} Collaborative Knowledge-aware Attentive Network for Recommender Systems}. In \bibinfo{booktitle}{\emph{Proceedings of the 43rd International {ACM} {SIGIR} conference on research and development in Information Retrieval, {SIGIR} 2020, Virtual Event, China, July 25-30, 2020}}, \bibfield{editor}{\bibinfo{person}{Jimmy~X. Huang}, \bibinfo{person}{Yi~Chang}, \bibinfo{person}{Xueqi Cheng}, \bibinfo{person}{Jaap Kamps}, \bibinfo{person}{Vanessa Murdock}, \bibinfo{person}{Ji{-}Rong Wen}, {and} \bibinfo{person}{Yiqun Liu}} (Eds.). \bibinfo{publisher}{{ACM}}, \bibinfo{pages}{219--228}.
\newblock
\urldef\tempurl%
\url{https://doi.org/10.1145/3397271.3401141}
\showDOI{\tempurl}


\bibitem[\protect\citeauthoryear{Wang, Zhang, Feng, and Chen}{Wang et~al\mbox{.}}{2014}]%
        {transH}
\bibfield{author}{\bibinfo{person}{Zhen Wang}, \bibinfo{person}{Jianwen Zhang}, \bibinfo{person}{Jianlin Feng}, {and} \bibinfo{person}{Zheng Chen}.} \bibinfo{year}{2014}\natexlab{}.
\newblock \showarticletitle{Knowledge Graph Embedding by Translating on Hyperplanes}. In \bibinfo{booktitle}{\emph{Proceedings of the Twenty-Eighth {AAAI} Conference on Artificial Intelligence, July 27 -31, 2014, Qu{\'{e}}bec City, Qu{\'{e}}bec, Canada}}, \bibfield{editor}{\bibinfo{person}{Carla~E. Brodley} {and} \bibinfo{person}{Peter Stone}} (Eds.). \bibinfo{publisher}{{AAAI} Press}, \bibinfo{pages}{1112--1119}.
\newblock


\bibitem[\protect\citeauthoryear{Xia, Huang, Xu, Dai, Zhang, Yang, Pei, and Bo}{Xia et~al\mbox{.}}{2021}]%
        {KHGT}
\bibfield{author}{\bibinfo{person}{Lianghao Xia}, \bibinfo{person}{Chao Huang}, \bibinfo{person}{Yong Xu}, \bibinfo{person}{Peng Dai}, \bibinfo{person}{Xiyue Zhang}, \bibinfo{person}{Hongsheng Yang}, \bibinfo{person}{Jian Pei}, {and} \bibinfo{person}{Liefeng Bo}.} \bibinfo{year}{2021}\natexlab{}.
\newblock \showarticletitle{Knowledge-Enhanced Hierarchical Graph Transformer Network for Multi-Behavior Recommendation}. In \bibinfo{booktitle}{\emph{Thirty-Fifth {AAAI} Conference on Artificial Intelligence, {AAAI} 2021, Thirty-Third Conference on Innovative Applications of Artificial Intelligence, {IAAI} 2021, The Eleventh Symposium on Educational Advances in Artificial Intelligence, {EAAI} 2021, Virtual Event, February 2-9, 2021}}. \bibinfo{publisher}{{AAAI} Press}, \bibinfo{pages}{4486--4493}.
\newblock


\bibitem[\protect\citeauthoryear{Xin, He, Zhang, Zhang, and Jose}{Xin et~al\mbox{.}}{2019}]%
        {RCF}
\bibfield{author}{\bibinfo{person}{Xin Xin}, \bibinfo{person}{Xiangnan He}, \bibinfo{person}{Yongfeng Zhang}, \bibinfo{person}{Yongdong Zhang}, {and} \bibinfo{person}{Joemon~M. Jose}.} \bibinfo{year}{2019}\natexlab{}.
\newblock \showarticletitle{Relational Collaborative Filtering: Modeling Multiple Item Relations for Recommendation}. In \bibinfo{booktitle}{\emph{Proceedings of the 42nd International {ACM} {SIGIR} Conference on Research and Development in Information Retrieval, {SIGIR} 2019, Paris, France, July 21-25, 2019}}, \bibfield{editor}{\bibinfo{person}{Benjamin Piwowarski}, \bibinfo{person}{Max Chevalier}, \bibinfo{person}{{\'{E}}ric Gaussier}, \bibinfo{person}{Yoelle Maarek}, \bibinfo{person}{Jian{-}Yun Nie}, {and} \bibinfo{person}{Falk Scholer}} (Eds.). \bibinfo{publisher}{{ACM}}, \bibinfo{pages}{125--134}.
\newblock
\urldef\tempurl%
\url{https://doi.org/10.1145/3331184.3331188}
\showDOI{\tempurl}


\bibitem[\protect\citeauthoryear{Yu, Ren, Sun, Gu, Sturt, Khandelwal, Norick, and Han}{Yu et~al\mbox{.}}{2014}]%
        {PER}
\bibfield{author}{\bibinfo{person}{Xiao Yu}, \bibinfo{person}{Xiang Ren}, \bibinfo{person}{Yizhou Sun}, \bibinfo{person}{Quanquan Gu}, \bibinfo{person}{Bradley Sturt}, \bibinfo{person}{Urvashi Khandelwal}, \bibinfo{person}{Brandon Norick}, {and} \bibinfo{person}{Jiawei Han}.} \bibinfo{year}{2014}\natexlab{}.
\newblock \showarticletitle{Personalized entity recommendation: a heterogeneous information network approach}. In \bibinfo{booktitle}{\emph{Seventh {ACM} International Conference on Web Search and Data Mining, {WSDM} 2014, New York, NY, USA, February 24-28, 2014}}, \bibfield{editor}{\bibinfo{person}{Ben Carterette}, \bibinfo{person}{Fernando Diaz}, \bibinfo{person}{Carlos Castillo}, {and} \bibinfo{person}{Donald Metzler}} (Eds.). \bibinfo{publisher}{{ACM}}, \bibinfo{pages}{283--292}.
\newblock
\urldef\tempurl%
\url{https://doi.org/10.1145/2556195.2556259}
\showDOI{\tempurl}


\bibitem[\protect\citeauthoryear{Zhang, Yuan, Lian, Xie, and Ma}{Zhang et~al\mbox{.}}{2016}]%
        {CKE}
\bibfield{author}{\bibinfo{person}{Fuzheng Zhang}, \bibinfo{person}{Nicholas~Jing Yuan}, \bibinfo{person}{Defu Lian}, \bibinfo{person}{Xing Xie}, {and} \bibinfo{person}{Wei{-}Ying Ma}.} \bibinfo{year}{2016}\natexlab{}.
\newblock \showarticletitle{Collaborative Knowledge Base Embedding for Recommender Systems}. In \bibinfo{booktitle}{\emph{Proceedings of the 22nd {ACM} {SIGKDD} International Conference on Knowledge Discovery and Data Mining, San Francisco, CA, USA, August 13-17, 2016}}, \bibfield{editor}{\bibinfo{person}{Balaji Krishnapuram}, \bibinfo{person}{Mohak Shah}, \bibinfo{person}{Alexander~J. Smola}, \bibinfo{person}{Charu~C. Aggarwal}, \bibinfo{person}{Dou Shen}, {and} \bibinfo{person}{Rajeev Rastogi}} (Eds.). \bibinfo{publisher}{{ACM}}, \bibinfo{pages}{353--362}.
\newblock
\urldef\tempurl%
\url{https://doi.org/10.1145/2939672.2939673}
\showDOI{\tempurl}


\bibitem[\protect\citeauthoryear{Zhao, Cheng, Zhu, Zheng, and Li}{Zhao et~al\mbox{.}}{2021}]%
        {UGRec}
\bibfield{author}{\bibinfo{person}{Xinxiao Zhao}, \bibinfo{person}{Zhiyong Cheng}, \bibinfo{person}{Lei Zhu}, \bibinfo{person}{Jiecai Zheng}, {and} \bibinfo{person}{Xueqing Li}.} \bibinfo{year}{2021}\natexlab{}.
\newblock \showarticletitle{UGRec: Modeling Directed and Undirected Relations for Recommendation}. In \bibinfo{booktitle}{\emph{{SIGIR} '21: The 44th International {ACM} {SIGIR} Conference on Research and Development in Information Retrieval, Virtual Event, Canada, July 11-15, 2021}}, \bibfield{editor}{\bibinfo{person}{Fernando Diaz}, \bibinfo{person}{Chirag Shah}, \bibinfo{person}{Torsten Suel}, \bibinfo{person}{Pablo Castells}, \bibinfo{person}{Rosie Jones}, {and} \bibinfo{person}{Tetsuya Sakai}} (Eds.). \bibinfo{publisher}{{ACM}}, \bibinfo{pages}{193--202}.
\newblock
\urldef\tempurl%
\url{https://doi.org/10.1145/3404835.3462835}
\showDOI{\tempurl}


\end{thebibliography}

\end{document}